\documentclass[preprint,review,12pt]{elsarticle}

\makeatletter
\long\def\pprintMaketitle{\clearpage
  \iflongmktitle\if@twocolumn\let\columnwidth=\textwidth\fi\fi
  \resetTitleCounters
  \def\baselinestretch{1}%
  \printFirstPageNotes
  \begin{\elsarticletitlealign}%
 \thispagestyle{pprintTitle}%
   \def\baselinestretch{1}%
    \Large\@title\par\vskip18pt%
    \ifx\@elsarticlenewpageafter\newpage@after@title
      \newpage
    \fi%
    \ifdoubleblind
      \vspace*{2pc}
    \else
      \normalsize\elsauthors\par\vskip10pt
      \footnotesize\itshape\elsaddress\par\vskip1em 
    \fi
    \ifx\@elsarticlenewpageafter\newpage@after@author
      \newpage
    \fi%
    \hrule\vskip12pt
    \ifvoid\absbox\else\unvbox\absbox\par\vskip10pt\fi
    \ifvoid\keybox\else\unvbox\keybox\par\vskip10pt\fi
    \hrule\vskip12pt
    \ifx\@elsarticlenewpageafter\newpage@after@abstract
      \newpage
    \fi%
    \end{\elsarticletitlealign}%
    \gdef\thefootnote{\arabic{footnote}}%
  }
\makeatother

\biboptions{numbers,sort&compress}

\usepackage{amsmath,amssymb,amsfonts}
\usepackage{graphicx}
\usepackage{textcomp}
\usepackage{xcolor}
\usepackage{color,soul}
\usepackage{multirow}
\usepackage{url}
\usepackage{caption}
\usepackage{algorithm,algorithmic}
\usepackage{enumitem}
\usepackage{subcaption}
\usepackage{amsmath, amssymb}
\usepackage{setspace}
\usepackage{longtable}

\usepackage[hidelinks]{hyperref}
\usepackage[nameinlink,noabbrev]{cleveref} 

\journal{Electric Power Systems Research}

\begin{document}

\begin{frontmatter}



\title{Impedance-Based VSC Unit Commitment with STATCOM Support under High IBG Penetration}


\author[KAUST]{Aoun Abbas}
\ead{aoun.abbas@kaust.edu.sa}
\author[TJU]{Zhongda Chu}
\author[KAUST]{Charalambos Konstantinou}
\cortext[cor1]{Corresponding author}

\affiliation[KAUST]{organization={King Abdullah University of Science and Technology},
            city={Thuwal},
            country={Kingdom of Saudi Arabia}}

\affiliation[TJU]{organization={Tianjin University},
            city={Tianjin},
            country={China}}

\begin{abstract}
The large-scale replacement of synchronous machines with inverter-based generation (IBG) introduces critical challenges to both voltage and frequency stability. This work builds on a mixed-integer second-order cone programming (MISOCP) framework that co-optimizes unit commitment (UC) model which embeds frequency-nadir constraints through synthetic inertia (SI) dispatch and an SOC voltage stability boundary for IBG buses. The formulation extends by modeling a STATCOM as a reactive-power decision variable in the same MISOCP model. A modified IEEE 30-bus system is used to assess three scheduling strategies: \textit{(i)} baseline UC with SI only, \textit{(ii)} voltage-stability-constrained (VSC) UC with SI, and \textit{(iii)} the joint UC with SI and reactive power support from IBGs. The impact of incorporating a 30~MVAr STATCOM at a weak grid location near the IBG buses is investigated. Simulation results show that the proposed framework enhances voltage security, maintains frequency-nadir compliance, and reduces operating cost, while STATCOM integration further improves dispatch feasibility under high IBG.
\end{abstract}




\end{frontmatter}

\section*{Nomenclature}
\vspace{-2mm}
{\small
\renewcommand{\arraystretch}{1.0}
\setlength{\tabcolsep}{4pt}

\begin{longtable}{p{0.22\linewidth} p{0.74\linewidth}}

\multicolumn{2}{l}{\textbf{Sets and indices}} \\[1mm]
$\mathcal{N}$ & Set of buses \\
$\mathcal{L}$ & Set of transmission lines \\
$\mathcal{G}_c$ & Set of synchronous generators (SGs) \\
$\mathcal{G}_v$ & Set of grid-forming units (GFMs/VSGs) \\
$\mathcal{C}$ & Set of grid-following IBGs \\
$g,c$ & Generator and IBG indices \\
$i,j$ & Bus indices \\
$t$ & Time index \\
$n$ & Scenario/node index \\[1mm]


\multicolumn{2}{l}{\textbf{Network and voltage quantities}} \\[1mm]
$Y$ & Bus admittance matrix \\
$Y^0$ & Network admittance from transmission lines \\
$Y^g$ & Admittance from SGs and GFMs \\
$Z$ & Bus impedance matrix, $Z = Y^{-1}$ \\
$V_i$ & Voltage at bus $i$ \\
$c_{ii}$ & Squared voltage magnitude at bus $i$ \\
$G_{ii}, B_{ii}$ & Self-conductance and susceptance at bus $i$ \\[1mm]

\multicolumn{2}{l}{\textbf{Commitment and dispatch variables}} \\[1mm]
$x_{g,t}$ & Commitment status of generator $g$ at time $t$ \\
$u_{g,t}$ & Start-up variable \\
$v_{g,t}$ & Shut-down variable \\
$P_g, Q_g$ & Active and reactive power of SGs \\
$P_c, Q_c$ & Active and reactive power of IBGs \\
$Q^{\mathrm{stat}}$ & Reactive power from STATCOM \\
$P^s$ & Load shedding \\[1mm]

\multicolumn{2}{l}{\textbf{Voltage stability quantities}} \\[1mm]
$\hat{P}_c,\hat{Q}_c$ & Equivalent injections at IBG bus $c$ \\
$\Gamma_c$ & Voltage stability margin at IBG bus $c$ \\
$z_c^{c'}$ & Mutual impedance ratio between IBG buses \\
$z_c^{1}$ & Self-impedance ratio \\
$\eta_m$ & Interaction term in the regression model \\[1mm]

\multicolumn{2}{l}{\textbf{Frequency security quantities}} \\[1mm]
$H$ & Aggregate system inertia \\
$H_{sj}$ & Synthetic inertia from IBG $j$ \\
$R$ & Primary frequency response coefficient \\
$\Delta P_L$ & Largest power disturbance \\
$\Delta f_{\mathrm{lim}}$ & Frequency deviation limit \\
$T_d$ & Response delivery time \\
$D$ & Load damping coefficient \\[1mm]

\multicolumn{2}{l}{\textbf{STATCOM and economic quantities}} \\[1mm]
$\bar Q_{\mathrm{stat}}$ & STATCOM reactive power rating \\
$I_{\max}$ & Maximum STATCOM current \\
$\pi(n)$ & Scenario probability \\
$C_g(n)$ & Generation cost at node $n$ \\
$\Delta t(n)$ & Duration of node $n$ \\
$S$ & Device rating in TCO analysis \\
$\ell_{\mathrm{stat}}, \ell_{\mathrm{sc}}$ & Device loss factors \\
$C_{\mathrm{loss}}$ & Annual loss cost \\
$C_{\mathrm{O\&M}}$ & Operation and maintenance cost \\

\end{longtable}
}

\setcounter{table}{0}

\section{Introduction}
\label{sec:1_Introduction}

The ongoing transition toward low-carbon power systems has significantly increased the penetration of renewable energy sources (RES), particularly wind and solar. Unlike synchronous generators (SGs), inverter-based generators (IBGs) do not inherently provide inertia, voltage regulation, or short-circuit support \cite{milano2018foundations}. Grid-following (GFL) IBGs, in particular, behave as controlled current sources with limited fault current capability, making them more vulnerable to voltage collapse under weak grid conditions \cite{modarresi2016review}. When multiple IBGs operate in close electrical proximity, their interactions can further degrade system strength, intensifying voltage instability risks \cite{wu2018scr}. At the same time, the displacement of SGs reduces available inertia. As a result, synthetic inertia (SI) from inverter controls has emerged as a practical means of mitigating frequency excursions \cite{teng2020synthetic, MEJIARUIZ2025110953}.

Traditional voltage stability studies have mainly focused on vulnerable load (PQ) buses far from SGs, assuming that generator buses (PV) are stabilized by SG excitation systems. However, with IBG-dominated generation, this assumption is no longer valid, and new stability indicators based on impedance ratios and system strength have been proposed \cite{chu2021short}. Beyond voltage-margin formulations, recent studies examine short-circuit current (SCC) and its impact on scheduling. In \cite{chutengSCC} assumptions about IBR over-current capability change the system-level SCC and the outcomes of UC problems include explicit SCC limits. This study focuses on static voltage margins. The static synchronous compensator (STATCOM) is modeled as a shunt VAr device and does not increase SCC in the adopted formulation. For this reason, SCC-constrained scheduling is treated as a complementary extension that can be added when protection limits become binding.

These coupled issues highlight the need for coordinated treatment of voltage and frequency stability under high IBG penetration. A significant advance was made in \cite{chu2023voltage}, which derived an analytical voltage stability criterion for IBG buses based on the system impedance matrix. This approach captures the interaction among multiple IBGs and embeds the resulting voltage-stability limits within the scheduling model. In \cite{chu2023voltage}, synchronous condensers (SCs) were used to enhance voltage stability under weak-grid conditions. However, SC deployment can involve high capital cost, longer installation time, and increased short-circuit levels, which may require protection upgrades.

As an alternative, STATCOMs provide fast-acting reactive support without increasing fault current or inertia requirements \cite{Alajrash2024EnergyReports}. Most prior studies treat STATCOMs in planning, optimal power flow (OPF), or dynamic simulation, rather than as explicit decision variables in UC. In \cite{FACTS_SCUC}, flexible AC transmission system (FACTS) devices are incorporated in security-constrained UC, and \cite{freq-nadir-UC} includes a frequency-nadir constraint in UC. However, these works do not consider a STATCOM together with impedance-based voltage-stability constraints at IBG buses. Independent planning studies and utility reports also show that STATCOMs are modular, relatively fast to deploy, and effective for weak-grid voltage support \cite{ISO-NE2021Stakeholder}.

To position the proposed formulation with respect to the most closely related scheduling models, Table~\ref{tab:qualitative_comparison} summarizes the key differences between this work and representative studies in the literature. As shown in Table~\ref{tab:qualitative_comparison}, the proposed method uniquely combines impedance-based voltage-stability constraints, frequency-security requirements, and explicit STATCOM-based reactive support within a unified MISOCP unit commitment framework.

\begin{table}[t]
\renewcommand{\arraystretch}{0.85}
\centering
\caption{Qualitative comparison with closely related UC formulations.}
\vspace{-3mm}
\label{tab:qualitative_comparison}
\setlength{\tabcolsep}{3.5pt}
\resizebox{\columnwidth}{!}{
\begin{tabular}{||c|c|c|c|c|c||}
\hline \hline
\textbf{Method} & \textbf{VSC} & \textbf{Frequency-nadir} & \textbf{Reactive} & \textbf{Device} & \textbf{MISOCP} \\
                & \textbf{cone} & \textbf{constraint}     & \textbf{support}  & \textbf{type}   & \textbf{} \\ \hline \hline

{\cite{chu2021short}}  & No  & Yes & No  & SG + IBG (SCC)      & No  \\ \hline
{\cite{chu2023voltage}}  & Yes & Yes & Yes & IBG + sync support  & Yes \\ \hline
{\cite{FACTS_SCUC}}  & No  & No  & Yes & FACTS (STATCOM)     & No  \\ \hline
{\cite{freq-nadir-UC}} & No  & Yes & No  & SG + RES + BESS     & No  \\ \hline
This work & Yes & Yes & Yes & STATCOM + IBG + SG & Yes \\ \hline \hline

\end{tabular}
}
\vspace{-4mm}
\end{table}

Motivated by these practical considerations, this work extends the impedance-based voltage stability constrained unit commitment (VSC-UC) framework in \cite{chu2023voltage} by incorporating a STATCOM as an explicit reactive-power decision variable in the same MISOCP model. The voltage-stability cone, the regression-based linearization, and the frequency-nadir constraint are retained from \cite{chu2023voltage}, while tractable convex relaxations are used for AC power flow \cite{kocuk2016strong} and stochastic scheduling under uncertainty \cite{teng2016stochastic}. The added STATCOM support is co-optimized with generator commitment, IBG dispatch, and synthetic inertia provision within this unified framework. In the modified IEEE 30-bus system, the STATCOM is placed at Bus~22, which is electrically close to the weak IBG buses and provides effective local reactive support.

The main contributions of this work are as follows:
\begin{itemize}[leftmargin=*,itemsep=1pt,topsep=2pt]
    \item The existing impedance-based VSC-UC formulation is extended by modeling a STATCOM as an explicit reactive-power decision variable within the same MISOCP framework.
    \item The STATCOM is represented using symmetric reactive-power limits and a voltage-dependent current constraint in rotated SOC form, while its effect is introduced only through the reactive nodal balance.
    \item The operational value of STATCOM support is quantified in a modified IEEE 30-bus system through a comparison of three cases: Base+SI, Case~I (VSC+SI), and Case~II (VSC+Q+SI).
    \item The scalability of the proposed framework is further examined on the IEEE 118-bus system to assess whether the same scheduling behavior is preserved in a larger network with higher renewable penetration.
\end{itemize}

This work does not aim to propose a new UC algorithm or improve computational efficiency. Its purpose is to extend an existing voltage-stability-constrained scheduling framework so that fast-acting shunt reactive support can be evaluated within the UC itself, rather than through a separate planning or OPF stage.

The remainder of this paper is organized as follows. Section \ref{sec:2_Voltage_stability_analysis} presents the voltage-stability analysis at IBG buses. Section \ref{sec:3_SOC_reformulation} describes the SOC reformulation and regression-based linearization. Section \ref{sec:4_UC_formulation} presents the VSC-UC model with STATCOM integration. Section \ref{sec:5_Case_study} reports the case studies, and Section \ref{sec:6_Conclusions} concludes the paper.

\section{Voltage Stability Analysis}
\label{sec:2_Voltage_stability_analysis}

We adopt the static voltage stability framework introduced in \cite{chu2023voltage}, extending it to our test system with modified generator configurations. Generators are categorized as synchronous machines $g_c \in \mathcal{G}_c$, grid-forming (GFM) virtual synchronous generators (VSG) $g_v \in \mathcal{G}_v$, and grid-following (GFL) IBGs $c \in \mathcal{C}$. The complete set of controllable voltage-regulating generators is defined as $\mathcal{G} = \mathcal{G}_c \cup \mathcal{G}_v$. Grid-following IBGs are treated separately as $c \in \mathcal{C}$, since they are modeled as current-injecting devices and do not contribute to the network admittance in the same way as SGs or GFMs.
The mappings $\Psi(g)$ and $\Phi(c)$ assign each generator or IBG to its corresponding bus index. For clarity, the mappings $\Psi(\cdot)$ and $\Phi(\cdot)$ are used solely to denote the bus indices associated with generators and IBGs, respectively. When appearing as subscripts of network matrices (e.g., $Z_{\Phi(c)\Phi(c')}$), they simply indicate the corresponding row and column of the matrix and do not imply any functional composition.
Let $\mathcal{N}$ denote the set of all buses in the transmission network. Generator buses and IBG buses are identified through $\Psi(g)$ and $\Phi(c)$. Load buses are the remaining elements of $\mathcal{N}$ and are modeled as passive nodes. Static voltage stability here refers to steady-state operation with acceptable voltage magnitudes under normal conditions and small disturbances. Transient voltage dynamics are not included. SGs and GFMs are treated as voltage-regulated buses, while GFL IBGs inject currents and are more susceptible under weak-grid conditions.
\vspace{-3mm}
\subsection{Jacobian-Based Derivation}
The fundamental power flow relation is expressed as $\mathbf{V} = \mathbf{Z} \cdot \mathbf{I}$, where $\mathbf{V}$ and $\mathbf{I}$ are bus voltage and current injection vectors, and $\mathbf{Z}$ is the system impedance matrix. For an IBG located at bus $\Phi(c)$, the corresponding row of $\mathbf{Z}$ yields:
\begin{equation}
V_{\Phi(c)} = \sum_{g \in \mathcal{G}} Z_{\Phi(c)\Psi(g)} I_g + \sum_{c' \in \mathcal{C}} Z_{\Phi(c)\Phi(c')} I_{c'}
\label{eq.2}
\end{equation}

\begin{figure}[ht]
    \centering
    \includegraphics[width=0.5\textwidth]{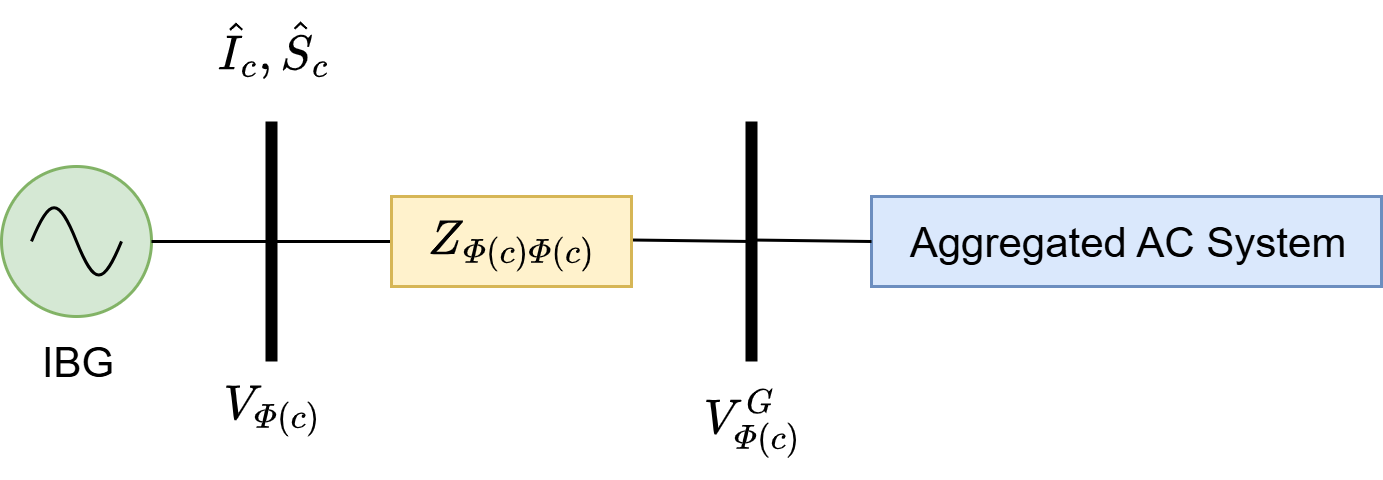}
    \vspace{-2mm}
    \caption{Equivalent two-bus representation for IBG bus $\Phi(c)$ and its aggregated network. }
    \vspace{-3mm}
    \label{fig:equivalent_ibg}
\end{figure}

Equation \eqref{eq.2} is the $\Phi(c)$-th row of $\mathbf{V}=\mathbf{Z}\mathbf{I}$, written by separating current injections from generators and grid-following IBGs. Load-only buses do not appear explicitly in the summations because they are modeled as passive elements without independent current injection. Their impact is captured implicitly in $\mathbf{Z}=\mathbf{Y}^{-1}$ through the network topology and line parameters. Fig.~\ref{fig:equivalent_ibg} shows the equivalent two-bus reduction introduced in \cite{chu2023voltage}, which can be written as:

\begin{equation}
V_{\Phi(c)} = V^{G}_{\Phi(c)} + Z_{\Phi(c)\Phi(c)} \hat{I}_c
\end{equation}
where $V^{G}_{\Phi(c)}$ represents the equivalent grid bus voltage and $\hat{I}_c$ the IBG current injection. The detailed Jacobian-based derivation and the expressions for the equivalent power injections can be found in \cite{chu2023voltage}. In this work, we adopt the resulting voltage-stability boundary directly for use in the UC formulation:
\begin{equation}
\hat{P}_c^2 + \hat{Q}_c^2 \le \big(\hat{Q}_c + \Gamma_c\big)^2
\label{eq:stability}
\end{equation}
\begin{equation}
\Gamma_c = \frac{|V^{G}_{\Phi(c)}|^2}{2\,|Z_{\Phi(c)\Phi(c)}|}
\label{eq:gamma}
\end{equation}

The equivalent power injections \((\hat P_c,\hat Q_c)\), along with their dependence on the network configuration, are approximated through learned impedance-ratio mappings as described in Section \ref{sec:3_SOC_reformulation}, following \cite{chu2023voltage}. Eqs. \eqref{eq:stability}-\eqref{eq:gamma} show that higher short-circuit capacity (smaller $|Z|$) improves the admissible active power injection, while additional reactive power increases stability margins. 
These impedance-ratio mappings depend on generator commitment decisions and inverter operating points, and include interaction terms that capture the combined effect of multiple devices on the local network impedance. The construction and interpretation of these interaction terms are described in detail in Section \ref{sec:3_SOC_reformulation}.

\section{SOC Reformulation of Voltage Stability Constraints}
\label{sec:3_SOC_reformulation}

The nonlinear voltage stability condition in \eqref{eq:stability} can be reformulated as a second-order cone (SOC) constraint involving $\hat{P}_c$, $\hat{Q}_c$, and $\Gamma_c$.
However, its components remain nonlinear because both $\hat{P}_c$ and $\hat{Q}_c$ depend on the network impedance matrix $Z$ and equivalent voltages, which vary with generator commitments. To embed these stability conditions into the optimization framework, we follow the SOC reformulation introduced in \cite{chu2023voltage}, which results in a voltage-stability constraint that can be directly enforced within the unit commitment formulation. This reformulation enables the voltage stability boundary to be enforced within a convex optimization framework, making it compatible with MISOCP-based unit commitment models.
\vspace{-3mm}
\subsection{Approximations for Equivalent Power Injections}

To handle the nonlinearity in the equivalent power injections, two standard approximations are adopted:
\textit{(i)} transmission lines exhibit small $R/X$ ratios, allowing resistances in $Z$ to be neglected, and
\textit{(ii)} voltage magnitudes at IBG buses remain close to one another, i.e., $\frac{|V_{\Phi(c)}|}{|V_{\Phi(c')}|} \approx 1$ under normal operating conditions \cite{wu2019method}. 
Hence, the equivalent injections at bus $\Phi(c)$ are:\vspace{-3mm}
\begin{align}
\hat{P}_c &= P_c + \sum_{\substack{c' \in \mathcal{C} \\ c' \neq c}} \frac{|Z_{\Phi(c)\Phi(c')}|}{|Z_{\Phi(c)\Phi(c)}|} P_{c'} \label{eq.6} \\
\hat{Q}_c &= Q_c + \sum_{\substack{c' \in \mathcal{C} \\ c' \neq c}} \frac{|Z_{\Phi(c)\Phi(c')}|}{|Z_{\Phi(c)\Phi(c)}|} Q_{c'} \label{eq.7}
\end{align}
and the local stability margin simplifies to:\vspace{-3mm}
\begin{equation}
\Gamma_c = \frac{1}{2|Z_{\Phi(c)\Phi(c)}|}
\end{equation}
These expressions remain nonlinear because the impedance matrix $Z$ depends on the commitment status of synchronous generators and the operating levels of grid-forming units. We therefore adopt a regression-based linearization discussed in \cite{chu2023voltage}.
In \eqref{eq.6} and \eqref{eq.7}, $P_c$ and $Q_c$ denote the local active and reactive power injections of IBG $c$ at its terminal bus $\Phi(c)$. The contributions of synchronous generators and grid-forming units are implicitly captured through the equivalent grid voltage and impedance, following the two-bus reduction in Fig. \ref{fig:equivalent_ibg}. As a result, the equivalent injections $\hat{P}_c$ and $\hat{Q}_c$ are expressed with respect to a single IBG, and no explicit summation over $g \in \mathcal{G}$ appears.
\vspace{-3mm}
\subsection{Formulation of Impedance Matrix and Linearization}
The system impedance matrix is obtained as:\vspace{-3mm}
\begin{align}
Z = Y^{-1}
\label{eq.9}\\
Y = Y^0 + Y^g
\end{align}

The inverse in \eqref{eq.9} is well-defined under normal operating conditions, assuming a connected transmission network with a designated reference bus, such that the reduced nodal admittance matrix is nonsingular. This is a standard assumption in steady-state power flow analysis and is satisfied for all operating points considered in this work.
Here, $Y^0$ represents transmission-line contributions, while $Y^g$ reflects the admittances of SGs and VSGs depending on their operational status, and is calculated as:
\begin{equation}
Y^g_{ij} =
\begin{cases}
\frac{1}{X_{gc}} x_{gc} & \text{if } i=j \ \&\ \exists gc \in \mathcal{G}_c: i=\Psi(gc) \\
\frac{1}{X_{gv}} \alpha_{gv} & \text{if } i=j \ \&\ \exists gv \in \mathcal{G}_v: i=\Psi(gv) \\
0 & \text{otherwise.}
\end{cases}
\label{eq.11}
\end{equation}
Here $x_{gc}$ is a binary variable indicating SG commitment, and $\alpha_{gv}\in[0,1]$ denotes the fraction of available renewable power from VSGs. 
In \eqref{eq.11}, $X_{gc}$ and $X_{gv}$ denote the equivalent transient reactances of synchronous generators and grid-forming virtual synchronous generators, respectively. These units are modeled as voltage-regulated sources and therefore contribute to the diagonal elements of the network admittance matrix. In contrast, grid-following IBGs are modeled as current-controlled devices and do not impose a voltage behind an internal reactance. As a result, their effect on the network is represented through current injections rather than admittance contributions, and they are not included in $Y^g$.
This modeling choice follows the impedance-based voltage stability framework adopted in~\cite{chu2023voltage}.

We model the on/off status of each synchronous generator $g \in \mathcal{G}$ at hour $t \in \mathcal{T}$ using a binary commitment variable $x_{g,t} \in \{0,1\}$. To capture unit transitions, we introduce binary start-up and shut-down variables $u_{g,t}, v_{g,t} \in \{0,1\}$ and enforce the standard commitment as:\vspace{-3mm}
\begin{equation}
x_{g,t} - x_{g,t-1} = u_{g,t} - v_{g,t}, \qquad \forall g \in \mathcal{G},\; \forall t \in \mathcal{T}.
\label{eq:uc_transition}
\end{equation}
Minimum up/down time constraints are imposed using $u_{g,t}$ and $v_{g,t}$ to prevent unrealistic rapid cycling. Direct symbolic expressions for $Z$ in terms of $\{x_{gc}, \alpha_{gv}\}$ are not tractable for realistic systems. Following \cite{chu2023voltage}, we approximate required impedance ratios with linear models. The two ratios of interest are: \vspace{-3mm}
\[
z_c^{c'} = \frac{|Z_{\Phi(c)\Phi(c')}|}{|Z_{\Phi(c)\Phi(c)}|}, \quad
z_c^1 = \frac{1}{|Z_{\Phi(c)\Phi(c)}|}
\]
Each ratio is expressed as:\vspace{-3mm}
\begin{equation}
z_c^\delta =
\sum_{gc \in \mathcal{G}_c} k_{c,gc}^\delta x_{gc}
+ \sum_{gv \in \mathcal{G}_v} k_{c,gv}^\delta \alpha_{gv}
+ \sum_{m \in \mathcal{M}} k_{c,m}^\delta \eta_m
\label{eq:15}
\end{equation}

where $\eta_m$ denote second-order interaction terms that capture the combined effect of multiple devices on the network impedance. These interaction terms are defined as pairwise products of decision variables, allowing the regression model to approximate the nonlinear dependence of impedance ratios on system configuration. Specifically, the interaction terms are constructed as:
\begin{equation}
\eta_m =
\begin{cases}
x_{gc} x_{gc'}, & \text{for } gc, gc' \in \mathcal{G}_c, \\
x_{gc} \alpha_{gv}, & \text{for } gc \in \mathcal{G}_c,\ gv \in \mathcal{G}_v, \\
\alpha_{gv} \alpha_{gv'}, & \text{for } gv, gv' \in \mathcal{G}_v,
\end{cases}
\end{equation}
for all interaction indices $m \in \mathcal{M}$. These second-order terms significantly improve approximation accuracy, while higher-order interactions are omitted to maintain computational tractability.
The index $\delta \in \{c', 1\}$ is introduced to represent both $z_c^{c'}$ and $z_c^{1}$ in a unified notation.
Coefficients with negligible magnitude are pruned to reduce model size without compromising accuracy. After an initial least-squares fit, coefficients below a small threshold are set to zero and the model is refit using only retained terms.
We verified that pruning reduces the number of interaction terms and associated auxiliary variables with minimal change in the error metrics in Table \ref{tab:maep}. The regression coefficients are obtained as: \vspace{-4mm}
\begin{equation}
\min_{\mathcal{K}^\delta} \sum_{\omega \in \Omega} \left(z_c^{\delta(\omega)} -
z_c^\delta|_{x_{gc}^{(\omega)},\alpha_{gv}^{(\omega)},\eta_m^{(\omega)}}\right)^2
\label{eq.14}
\end{equation}
with per-sample prediction as: \vspace{-4mm}
\begin{equation}
z_c^{\delta(\omega)} =
z_c^\delta \big|_{x_{gc}^{(\omega)},\alpha_{gv}^{(\omega)},\eta_m^{(\omega)}}
\end{equation}
where $\omega \in \Omega$ denotes a data sample in the offline dataset $\Omega$ and $n_v$ is the number of discretization levels used for the continuous VSG strength parameter $\alpha_{g_v}\in[0,1]$ for each $g_v \in \mathcal{G}_v$. The dataset size is: \vspace{-3mm}
\begin{equation}
|\Omega| = 2^{|\mathcal{G}_c|}\, n_v^{|\mathcal{G}_v|}.
\end{equation}

Each data sample $\omega \in \Omega$ corresponds to a synthetic steady-state operating condition defined by a synchronous generator commitment pattern $\mathbf{x}^{(\omega)}$ and a vector of grid-forming strength parameters $\boldsymbol{\alpha}^{(\omega)}$. For each sample, $\mathbf{Y}^{(\omega)}=\mathbf{Y}^0+\mathbf{Y}^{g,(\omega)}$ is constructed using \eqref{eq.11}, then $\mathbf{Z}^{(\omega)}=(\mathbf{Y}^{(\omega)})^{-1}$ is computed and the targets $z_c^\delta(\omega)$ are evaluated at the IBG buses of interest.
In this process, network topology and line parameters remain fixed, while variations in generator commitments and grid-forming strength parameters drive changes in the impedance matrix and corresponding Z-ratios. The dataset is generated entirely offline and is independent of the stochastic scenario tree used in the UC formulation, ensuring that the regression model captures only structural network effects rather than scenario-specific variations.
The regression in \eqref{eq.14} is also performed offline. The fitted coefficients $K^\delta$ are fixed parameters in the UC and are not decision variables.

Substituting \eqref{eq:15} into \eqref{eq.6} yields:\vspace{-3mm}
\begin{equation}
\hat{P}_c = P_c + \sum_{\substack{c' \in \mathcal{C}\\c'\neq c}}
\left(\sum_{gc\in\mathcal{G}_c} k^{c'}_{c,gc}x_{gc}
+ \sum_{gv\in\mathcal{G}_v} k^{c'}_{c,gv}\alpha_{gv} \right. \\
\left. + \sum_{m\in\mathcal{M}} k^{c'}_{c,m}\eta_m\right) P_{c'}, \qquad \forall c\in\mathcal{C}
\end{equation}
and $\hat{Q}_c$ follows the same form. Since $P_{c'}$ is continuous, cross-terms such as $x_{gc}P_{c'}$ and $\eta_mP_{c'}$ are bilinear. To maintain tractability, these bilinear terms are linearized using standard auxiliary variable techniques for binary-continuous products, allowing the formulation to remain within the MISOCP framework.

\subsection{Dataset and Regression Performance}
In this study, we focus on two grid-following IBGs located at buses 23 and 24 of the IEEE 30-bus test system, which are among the electrically weakest buses under high renewable penetration. To characterize the local and coupled voltage stability properties at these buses, four impedance-based Z-ratios are considered: the self-ratios $z^1_{23}$ and $z^1_{24}$, and the mutual ratios $z^{24}_{23}$ and $z^{23}_{24}$. The self-ratios capture the short-circuit strength seen at each IBG bus, while the mutual ratios quantify the coupling effect between the two IBGs through the network impedance. The resulting Z-ratios at buses 23 and 24 are used as regression targets for the linear surrogate models.

\begin{figure}[t]
    \centering
    \includegraphics[width=0.49\textwidth]{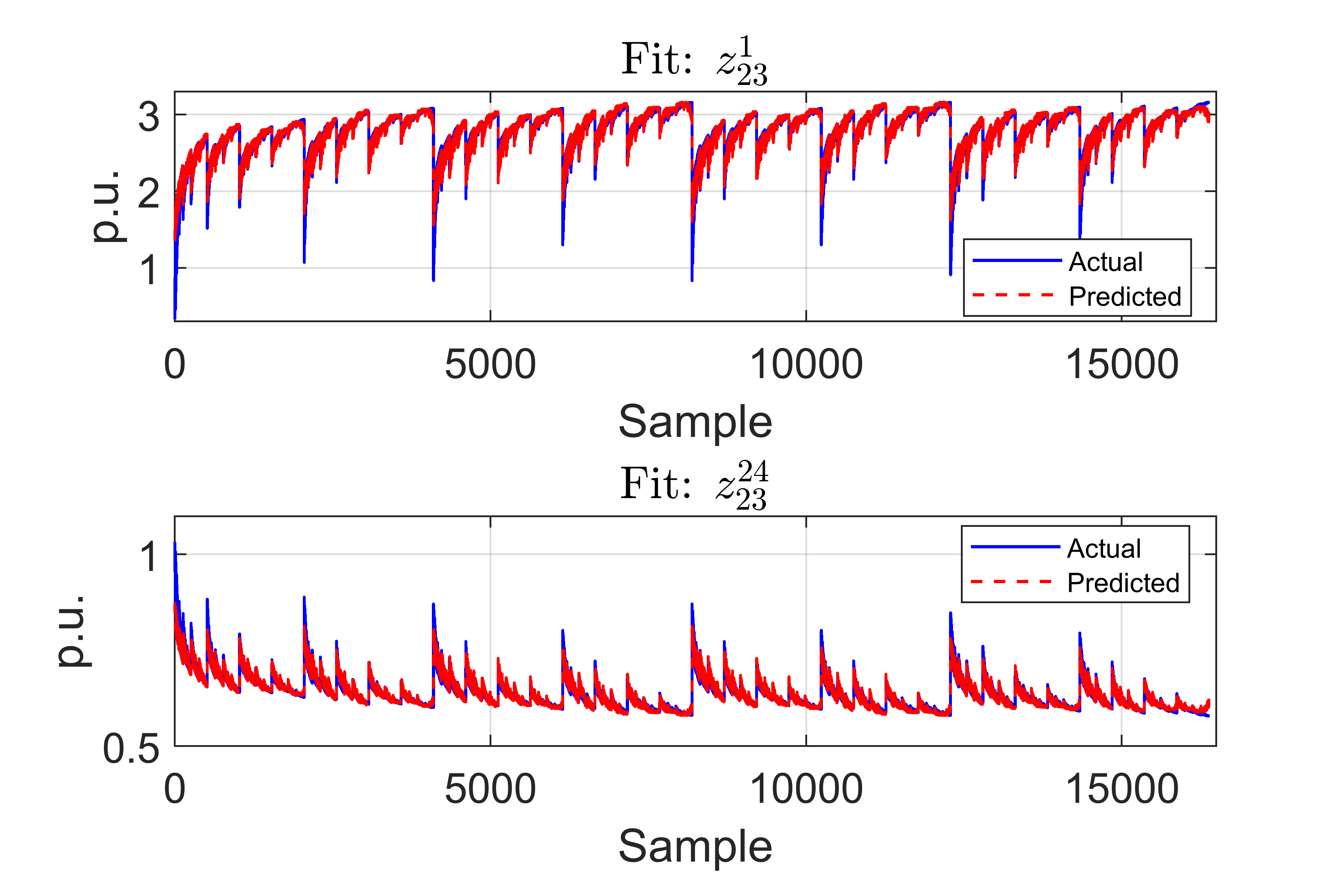}
    \hfill
    \includegraphics[width=0.49\textwidth]{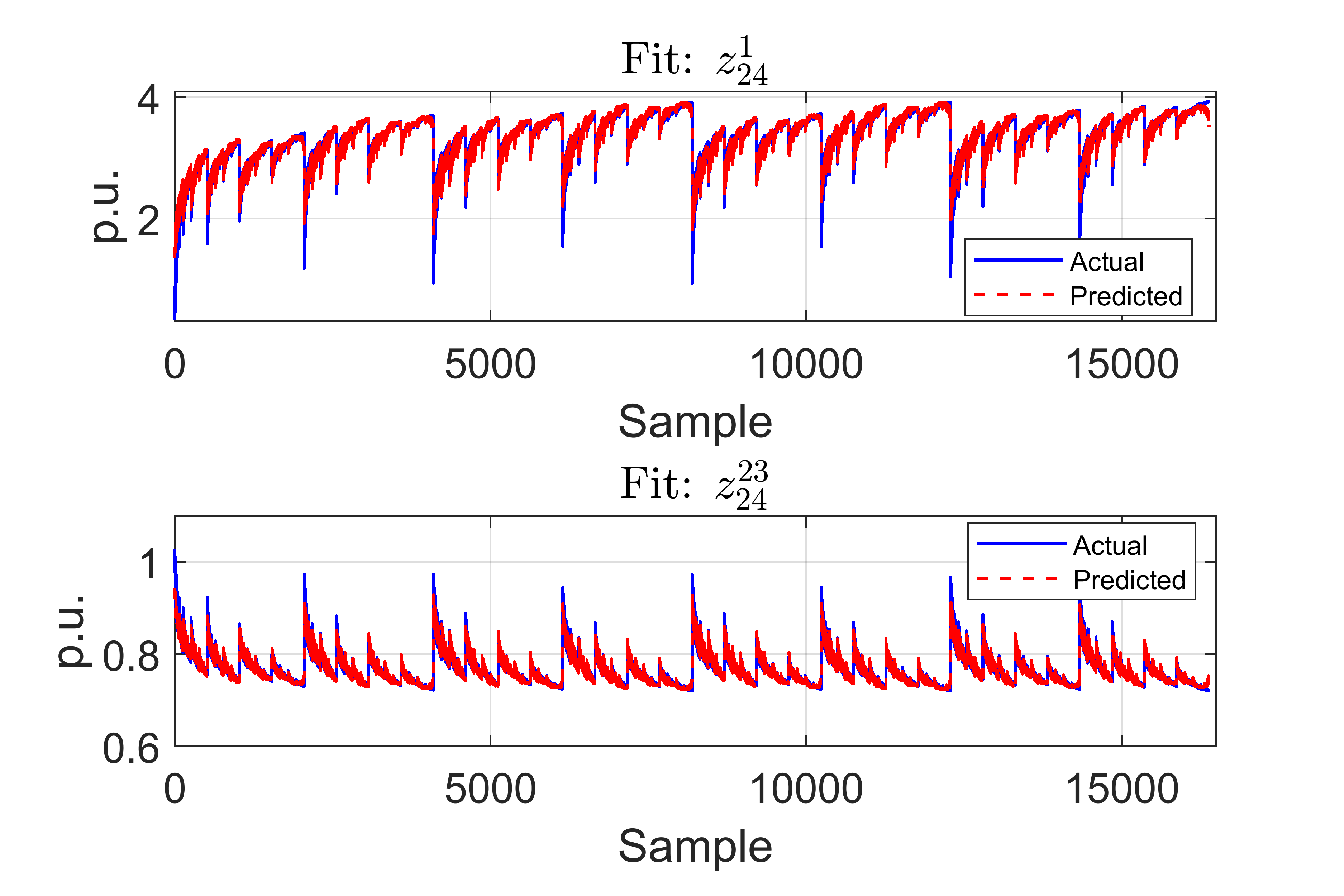}
    \vspace{-3mm}
    \caption{Regression performance: actual vs. predicted values across all samples for the four $Z$-ratio targets.}
    \vspace{-3mm}
    \label{fig:regression_performance}
\end{figure}

Fig.~\ref{fig:regression_performance} illustrates the variation of the four impedance-based Z-ratio targets
$\{z^{1}_{23},\,z^{24}_{23},\,z^{1}_{24},\,z^{23}_{24}\}$ across the generated dataset, while
Table~\ref{tab:maep} reports the corresponding regression accuracy metrics.
In the figure, the blue curves represent the actual Z-ratio values computed from the network impedance matrix, while the dashed red curves denote the predictions of the linear regression models. The self-ratio terms $z^{1}_{23}$ and $z^{1}_{24}$ exhibit wider variation, reflecting changes in local system strength at the IBG buses, whereas the mutual ratios $z^{24}_{23}$ and $z^{23}_{24}$ remain closer to unity but still respond to network configuration and dispatch conditions.
Across all four targets, the predicted values closely follow the actual series, with only small deviations at operating-point transitions.
The regression models are trained using 16{,}384 offline steady-state samples generated independently of the stochastic UC scenarios.
These results indicate that the selected feature set captures both local voltage strength and inter-IBG coupling effects with sufficient accuracy for use in the VSC-UC formulation.

\begin{figure}[t]
    \centering
    \includegraphics[width=0.5\textwidth]{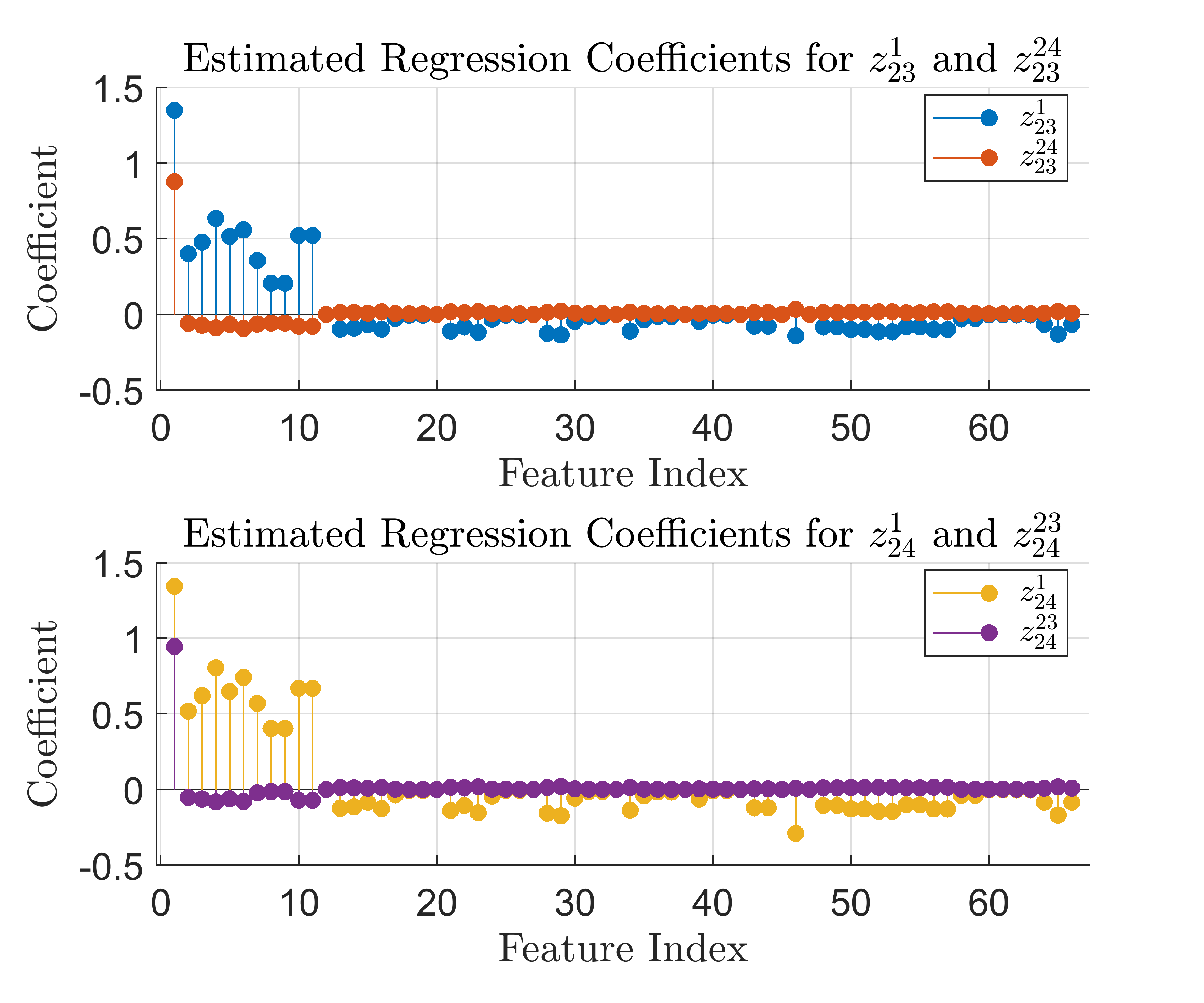}
    \vspace{-3mm}
    \caption{Estimated regression coefficients for each target.}
    \label{fig:coefficients}
    \vspace{-3mm}
\end{figure}

Fig.~\ref{fig:coefficients} shows the estimated regression coefficients for each Z-ratio target.
Most coefficients are close to zero, while a limited subset has dominant influence, indicating which commitment and interaction terms drive the impedance seen by the IBG buses.
This sparsity pattern supports the pruning step applied during model selection.
Table~\ref{tab:maep} reports MAEP values below 1.5\% for all targets, validating that the reduced linear models are accurate enough to embed within the MISOCP formulation.

\begin{table}[t]
\renewcommand{\arraystretch}{0.8} 
\centering
\caption{Regression accuracy summary: MSE and MAEP.}\vspace{-3mm}
\label{tab:maep}
\begin{tabular}{||c|c|c||}
\hline \hline
\textbf{Target} & \textbf{MSE (p.u.$^2$)} & \textbf{MAEP (\%)} \\ \hline \hline
$z^{1}_{23}$    & 0.003130 & 1.42 \\ \hline
$z^{24}_{23}$   & 0.000059 & 0.73 \\ \hline
$z^{1}_{24}$    & 0.004495 & 1.47 \\ \hline
$z^{23}_{24}$   & 0.000028 & 0.46 \\ \hline \hline
\end{tabular}
\vspace{-4mm}
\end{table}

\section{Voltage Stability-Constrained UC Formulation}
\label{sec:4_UC_formulation}

This section builds on the stochastic VSC-UC formulation in \cite{chu2023voltage}. The objective function, scenario-tree construction, rolling implementation, and frequency-nadir constraint follow that framework. The main extension introduced here is the explicit inclusion of STATCOM reactive support within the same MISOCP scheduling model. For completeness, the stochastic UC objective minimizes the expected operating cost over scenario-tree nodes $n \in \mathcal{N}$:
\begin{equation}
\min \sum_{n \in \mathcal{N}} \pi(n) \left( \sum_{g \in \mathcal{G}} C_g(n) + \Delta t(n)\, c_s\, P^s(n) \right)
\label{eq:obj}
\end{equation}
where $\pi(n)$ is the probability of node $n$, $C_g(n)$ includes startup, no-load, and marginal costs, and $c_s P^s(n)$ penalizes load shedding. The SUC is solved in a 24-hour rolling horizon, where only the first-hour decision is implemented and the scenario tree is rebuilt from updated forecasts. Standard UC constraints, including capacity limits, ramping limits, and minimum up/down times, are included in the model but are not repeated here because they follow conventional stochastic UC formulations; see \cite{teng2016stochastic}.

Uncertainty in load demand and wind generation is represented by a quantile-based scenario tree. At each hour, forecast errors are discretized into quantile bins and representative realizations are selected to form scenario paths over the look-ahead horizon. Node probabilities $\pi(n)$ are obtained from the selected quantile probabilities and normalized so that $\sum_{n\in\mathcal{N}}\pi(n)=1$. Commitment variables are first-stage decisions shared across scenarios, while dispatch, network, load shedding, and voltage-stability auxiliary variables are scenario-dependent recourse decisions. Scenario reduction is not applied, since preserving quantile-based extreme realizations is important for assessing voltage-stability constraints.
\vspace{-3mm}
\subsection{AC Power Flow and Power Balance Constraints}
As voltage security depends on reactive flows, a DC approximation is insufficient. We employ an SOC-relaxed AC network model following \cite{kocuk2016strong}. Here, $\mathcal{B}$ denotes the set of buses, $\mathcal{L}$ the set of transmission lines, and $\mathcal{L}(i)$ the set of buses connected to bus $i$. For each line $ij\in\mathcal{L}$, define:\vspace{-3mm}
\begin{align}
c_{ij} &= |V_i||V_j|\cos(\theta_i - \theta_j) \\
s_{ij} &= -|V_i||V_j|\sin(\theta_i - \theta_j) 
\end{align}
where $|V_i|$ and $\theta_i$ represent the voltage magnitude and phase angle at bus $i$. The nodal balances are:\vspace{-3mm}
\begin{align}
P^G_i - P^D_i &= G_{ii} c_{ii} + \sum_{j \in \mathcal{L}(i)} P_{ij}, \quad \forall i \in \mathcal{B} \\
Q^G_i - Q^D_i &= -B_{ii} c_{ii} + \sum_{j \in \mathcal{L}(i)} Q_{ij}, \quad \forall i \in \mathcal{B} \label{eq:23}
\end{align}
The parameters $G_{ii}$ and $B_{ii}$ denote the self-conductance and self-susceptance at bus $i$ derived from the network admittance matrix.
A STATCOM at bus $b_S$ (Bus~22 in our case study) is modeled by a reactive-power decision variable $Q^{\text{stat}}$ that enters only the reactive nodal balance at its bus:\vspace{-3mm}
\begin{equation}
Q^G_{b_S} - Q^D_{b_S} + Q^{\text{stat}} \;=\; -B_{b_S b_S} c_{b_S b_S} + \sum_{j \in \mathcal{L}(b_S)} Q_{b_S j}
\end{equation}
while \eqref{eq:23} remains unchanged for all $i \neq b_S$. The STATCOM provides reactive power only. Active power is fixed to zero and converter losses are neglected in the baseline experiments. The voltage box constraints and the symmetry and asymmetry conditions for auxiliary variables are:\vspace{-3mm}
\begin{align}
V^2_{\min,i} \leq c_{ii} \leq V^2_{\max,i}, \quad \forall i \in \mathcal{B}\\
c_{ij} = c_{ji},\quad s_{ij}=-s_{ji},\quad \forall ij\in\mathcal{L} \\
c_{ij}^2 + s_{ij}^2 \le c_{ii}c_{jj},\quad \forall ij\in\mathcal{L} \label{eq:26}
\end{align}

The rotated SOC in \eqref{eq:26} enforces a convex relaxation of the voltage-angle feasibility set. The above formulation adopts several modeling simplifications that should be stated explicitly. First, the STATCOM is represented as a shunt reactive-power device and therefore enters only the reactive nodal balance. It does not contribute active power, inertia, or short-circuit strength in the present formulation.
Allowing even a partial short-circuit contribution from the STATCOM could influence the resulting commitment and dispatch decisions under high IBR penetration, but this effect is not considered here and is left for future investigation.
Second, converter losses are neglected inside the UC model so that the scheduling problem remains focused on operational dispatch decisions. These assumptions are consistent with the intended role of the STATCOM in this work, which is to provide fast local voltage support rather than to modify system strength or frequency response. Their implications are discussed further in the case study section.

\subsection{IBG Capacity and STATCOM Capability Constraints}

For each grid-following IBG $c \in \mathcal{C}$, the voltage-stability constraint
derived in Section \ref{sec:3_SOC_reformulation} is enforced in SOC form as:\vspace{-3mm}
\begin{equation}
\hat{P}_c^2 + \hat{Q}_c^2 \le \left(\hat{Q}_c + \Gamma_c\right)^2,
\qquad \forall c \in \mathcal{C}.
\label{eq:vsc_soc}
\end{equation}
The equivalent injections $\hat P_c$ and $\hat Q_c$ are defined and linearized as described in Section \ref{sec:3_SOC_reformulation}. The IBG apparent-power capability is enforced by:\vspace{-3mm}
\begin{equation}
P_c^2 + Q_c^2 \;\le\; S_{c,\max}^2, \qquad \forall c\in\mathcal{C}
\label{eq.28}
\end{equation}
with $S_{c,\max}$ set to \(1.0\) p.u. following \cite{teng2020synthetic}, excluding short-term overloading. The local voltage-stability margin is then approximated using the learned self-impedance ratio model as:\vspace{-3mm}
\begin{equation}
\Gamma_c =
\frac{1}{2}
\left(
\sum_{gc \in \mathcal{G}_c} k^{1}_{c,gc} x_{gc}
+ \sum_{gv \in \mathcal{G}_v} k^{1}_{c,gv} \alpha_{gv}
+ \sum_{m \in \mathcal{M}} k^{1}_{c,m} \eta_m
\right),
\qquad \forall c \in \mathcal{C}.
\label{eq:29d}
\end{equation}
The STATCOM reactive injection is constrained by a symmetric box bound and a voltage-scaled current limit:\vspace{-3mm}
\begin{equation}
-\bar{Q}_{\text{stat}} \le Q^{\text{stat}} \le \bar{Q}_{\text{stat}}
\end{equation}
\begin{equation}
\left(Q^{\text{stat}}\right)^2 \le I_{\max}^2\, c_{b_S b_S}, 
\qquad c_{b_S b_S} = |V_{b_S}|^2
\end{equation}
which is imposed in rotated-SOC form. This capability model reflects the practical role of the STATCOM as a fast reactive-power source whose impact is local and voltage-dependent.

\subsection{Frequency Constraints}
To capture the effect of synchronous inertia and synthetic inertia on system frequency security, frequency-nadir constraints from \cite{teng2020synthetic} are included. The nadir behavior depends on the aggregate synchronous inertia $H$, the primary frequency response coefficient $R$, and the synthetic inertia contributions $H_{sj}$ provided by wind turbines.
Following the SOC reformulation in \cite{chu2023voltage}, the frequency-nadir condition is enforced as:\vspace{-3mm}
\begin{equation}
HR \geq
\underbrace{%
\frac{\Delta P_{L}^{2} T_{d}}{4 \Delta f_{\mathrm{lim}}}
- \frac{\Delta P_{L} T_{d} D}{4}%
}_{x_{1}^{2}}
+ \frac{\Delta P_{L} T_{d}}{4}
\sum_{j\in\mathcal{F}} \gamma_{j} H_{sj}^{2},
\label{eq:32}
\end{equation}
In \eqref{eq:32}, $\Delta P_L$ denotes the largest power disturbance, $\Delta f_{\mathrm{lim}}$ is the allowable frequency deviation, $T_d$ is the response delivery time, and $D$ represents load damping. The participation factor $\gamma_j$ indicates whether IBG $j$ contributes synthetic inertia. The auxiliary variable $x_1$ collects constant terms and is guaranteed to be real under the practical condition $\Delta P_L / \Delta f_{\mathrm{lim}} > D$. The decision variables in this constraint are $H$, $R$, and $H_{sj}$. 
This formulation makes the role of synthetic inertia explicit in the UC model and allows its effect on feasible scheduling decisions to be evaluated in the case studies.

\vspace{-3mm}
\section{Case Studies}
\label{sec:5_Case_study}
\vspace{-3mm}
To validate the proposed framework, we conduct numerical simulations on a modified IEEE 30-bus benchmark system. This section outlines the test system configuration and computational environment.

\subsection{System Setup and Computational Environment}

\begin{figure}[t]
    \centering
    \includegraphics[width=0.5\textwidth]{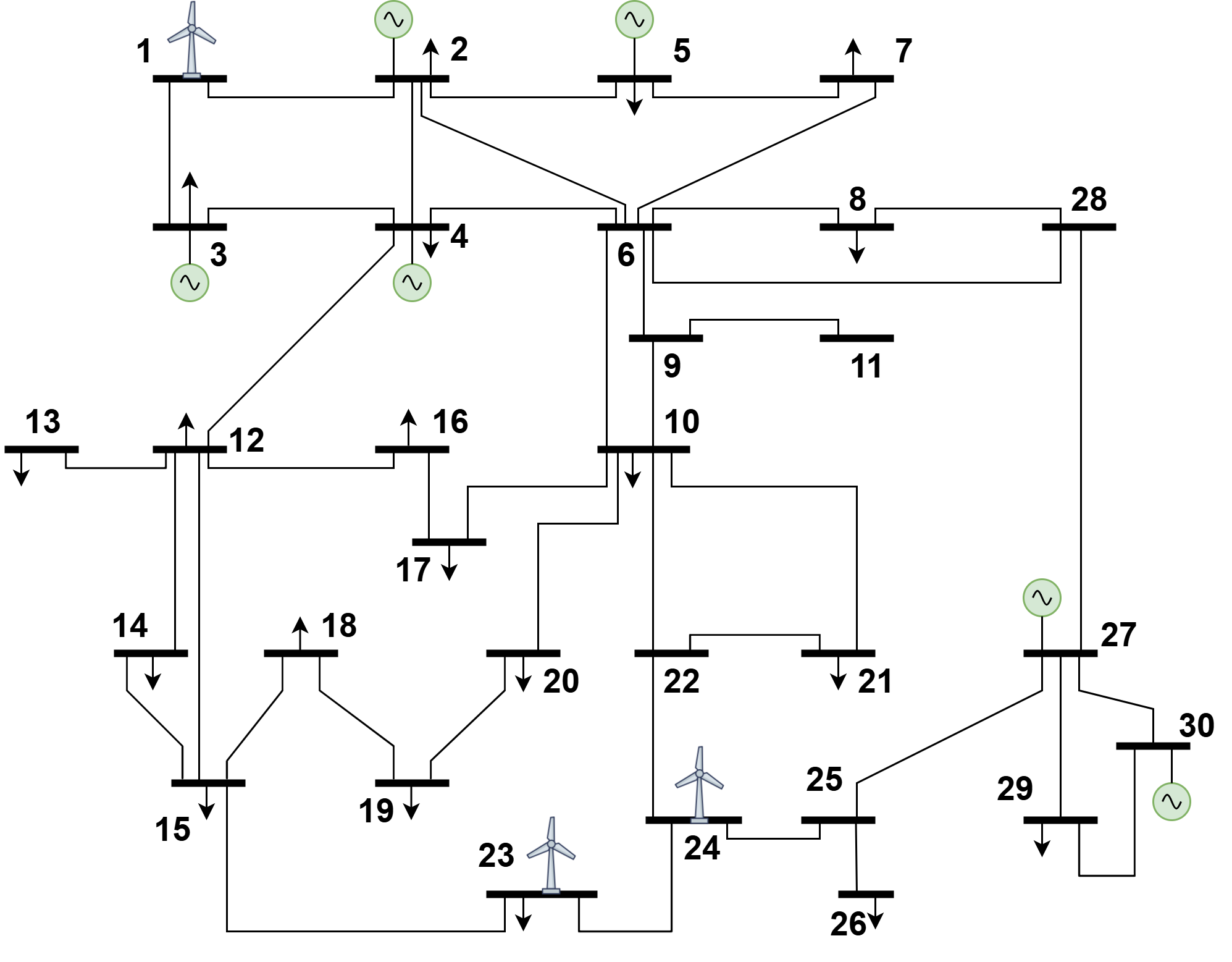} 
    \vspace{-3mm}
    \caption{Modified IEEE 30-bus test system with SGs, GFMs, and GFL-IBGs.}
    \vspace{-4mm}
    \label{fig:IEEE-30}
\end{figure}

The case study uses a modified IEEE 30-bus benchmark system designed to represent a high-renewable operating condition. IBGs (GFM and GFL) supply 62.5\% of the installed capacity and synchronous generation supplies 37.5\%. The network includes 41 transmission lines, 8 SGs, 2 GFMs, and 2 GFL-IBGs. The SGs are located at buses $\{2,2,3,4,5,27,30,30\}$, the GFMs are connected at bus~1, and the GFL-IBGs are connected at buses~23 and~24. The modified system layout is shown in Fig.~\ref{fig:IEEE-30}. The base power is $S_B=100$~MVA and the total demand varies in the range $P_D\in[160,410]$~MW. SG costs are modeled with quadratic functions and IBGs are assigned negligible marginal cost to represent RES. The maximum active power of each GFL-IBG at buses 23 and 24 is 300~MW, and the two GFMs at bus~1 provide up to 100~MW in total. All simulations are implemented in MATLAB R2024b using YALMIP and solved with the FICO Xpress optimizer. The model is formulated as MISOCP. Experiments are run on a Windows workstation with an AMD Ryzen~9~9950X 16-core CPU and 6.9~GB RAM.

\subsection{Impact of Wind Penetration on Voltage Stability}

\begin{figure}[ht]
    \centering
    \includegraphics[width=0.5\linewidth]{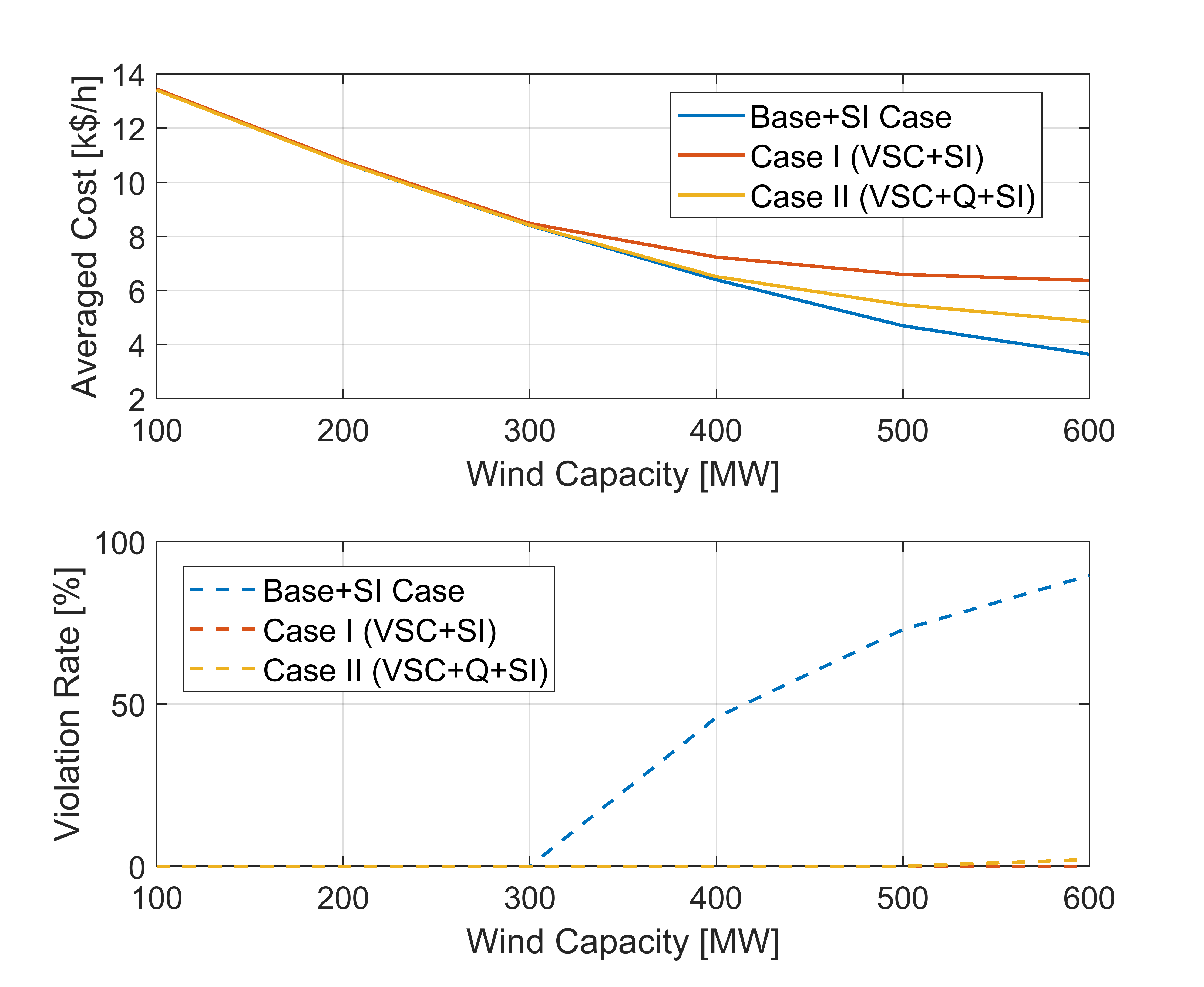}
    \vspace{-3mm}
    \caption{Impact of installed wind capacity on average operational cost and dashed lines represent the voltage violation rate.}
    \vspace{-6mm}
    \label{fig:cost_viol}
\end{figure}

Fig.~\ref{fig:cost_viol} shows the one-day ahead average operating cost and the voltage-stability violation rate as the installed wind capacity increases from 100~MW to 600~MW. 
Three cases are compared. In Base+SI, the unit commitment includes synthetic inertia but does not enforce the voltage-stability cone. In Case~I (VSC+SI), the cone is enforced while IBG reactive power is not optimized. In Case~II (VSC+Q+SI), both the cone and IBG reactive-power support are included.
The cost curves are close to one another up to about 300~MW, which indicates that voltage stability is not yet a limiting factor at lower wind penetration. Beyond this point, the differences become clear. Base+SI remains the cheapest because it does not enforce the voltage-stability constraint. Case~I (VSC+SI) becomes the most expensive, since the schedule must satisfy both the SOC voltage-stability condition and the AC network constraints without reactive support from the IBGs. Case~II (VSC+Q+SI) reduces this penalty by allowing reactive-power support from the IBGs, which enlarges the feasible operating region and leads to lower curtailment and lower operating cost than Case~I (VSC+SI) at higher wind levels.

The violation-rate curves show an even stronger distinction. Both Case~I (VSC+SI) and Case~II (VSC+Q+SI) maintain zero voltage-stability violations across the full wind range because the cone constraint is enforced directly in the optimization. In contrast, Base+SI remains violation-free only up to 300~MW, then rises sharply to about 45.8\% at 400~MW, 72.9\% at 500~MW, and 89.6\% at 600~MW. This confirms that voltage-stability constraints become necessary as renewable penetration increases, and that reactive-power support can recover much of the flexibility lost in the constrained schedule.
Across all three cases, the frequency-security constraints remain satisfied. In the tested wind range, the frequency-violation rate remains zero and the nadir and RoCoF slacks stay positive in all cases. This confirms that synthetic inertia is active in the scheduling model and that the differences observed in Fig.~\ref{fig:cost_viol} are driven primarily by voltage-stability enforcement rather than by loss of frequency security. Additional results on curtailment, voltage support, and frequency metrics are discussed next.

\subsection{Impact of STATCOM Rating on Operating Cost}

\begin{figure}[ht]
    \centering
    \includegraphics[width=0.5\linewidth]{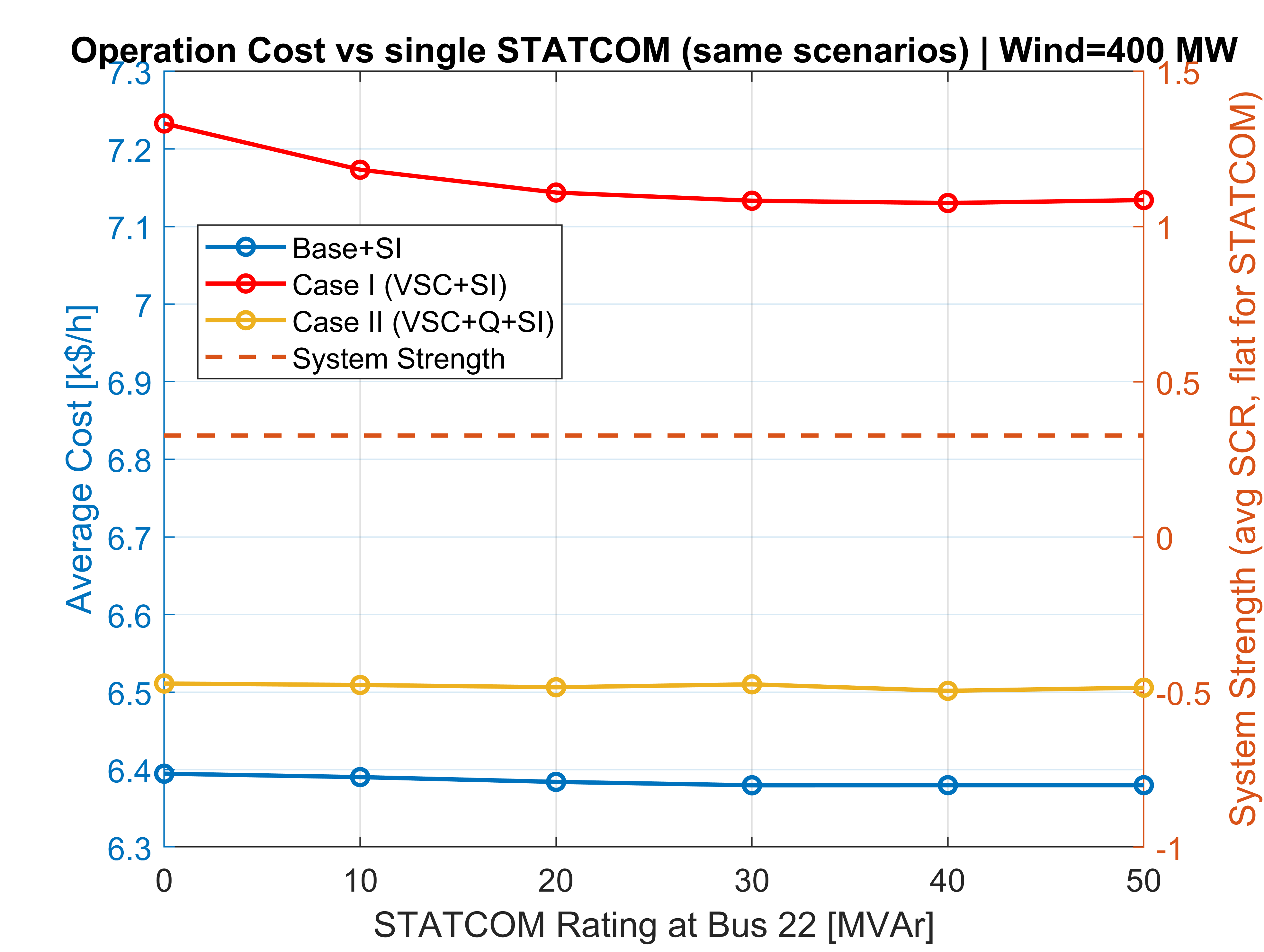}
    \vspace{-3mm}
    \caption{Impact of STATCOM rating on average operational cost at 400~MW and dashed line represents constant system-strength with STATCOM.}
    \vspace{-8mm}
    \label{fig:statcom_cost_bus22}
\end{figure}

To assess the effect of reactive-power reinforcement, a single STATCOM is placed at Bus~22 and its rating is varied over $\bar Q_{\text{stat}} \in \{0,10,20,30,40,50\}$~MVAr under the same one-day scenario set. Fig.~\ref{fig:statcom_cost_bus22} reports the resulting average operating cost at 400~MW wind for Base+SI, Case~I (VSC+SI), and Case~II (VSC+Q+SI). For reference, the dashed curve on the right axis shows a flat system-strength indicator, reflecting the fact that the STATCOM is modeled as a shunt reactive-power device and therefore does not increase short-circuit strength in the adopted formulation.

The most pronounced benefit appears in Case~I (VSC+SI). Its average operating cost decreases from about 7.23~k\$/h at 0~MVAr to about 7.13~k\$/h at 30~MVAr, after which the curve becomes nearly flat. This indicates that most of the attainable operating-cost benefit is captured by a 30~MVAr device, while larger ratings provide only marginal additional value. In contrast, Base+SI and Case~II (VSC+Q+SI) are much less sensitive to the STATCOM rating. Base+SI remains voltage-insecure because the voltage-stability cone is not enforced, whereas Case~II (VSC+Q+SI) already has access to IBG reactive-power support and therefore requires less additional support from the STATCOM. These results show that the STATCOM is most valuable when the system is forced to satisfy the voltage-stability constraint without reactive-power flexibility from the IBGs. For this reason, a 30~MVAr STATCOM at Bus~22 is adopted in the subsequent discussion as a practical and cost-effective reinforcement level.

\subsection{Additional Voltage, Curtailment, and Frequency Results}

\begin{figure}[ht]
    \centering
    \includegraphics[width=0.5\linewidth]{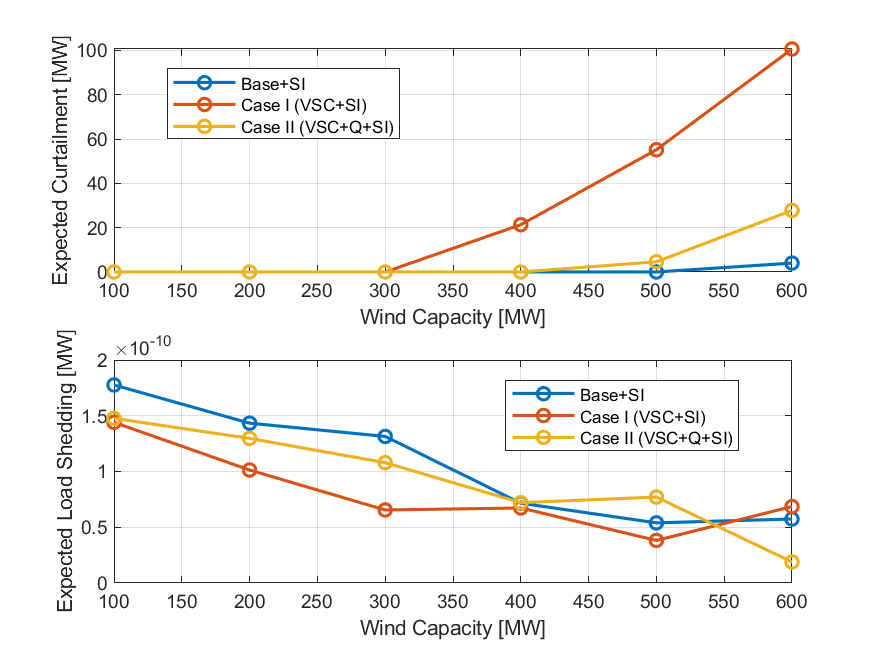}
    \vspace{-3mm}
    \caption{Expected wind curtailment and load shedding versus installed wind capacity.}
    \label{fig:curtail_shed}
\end{figure}

\begin{figure}[ht]
    \centering
    \includegraphics[width=0.5\linewidth]{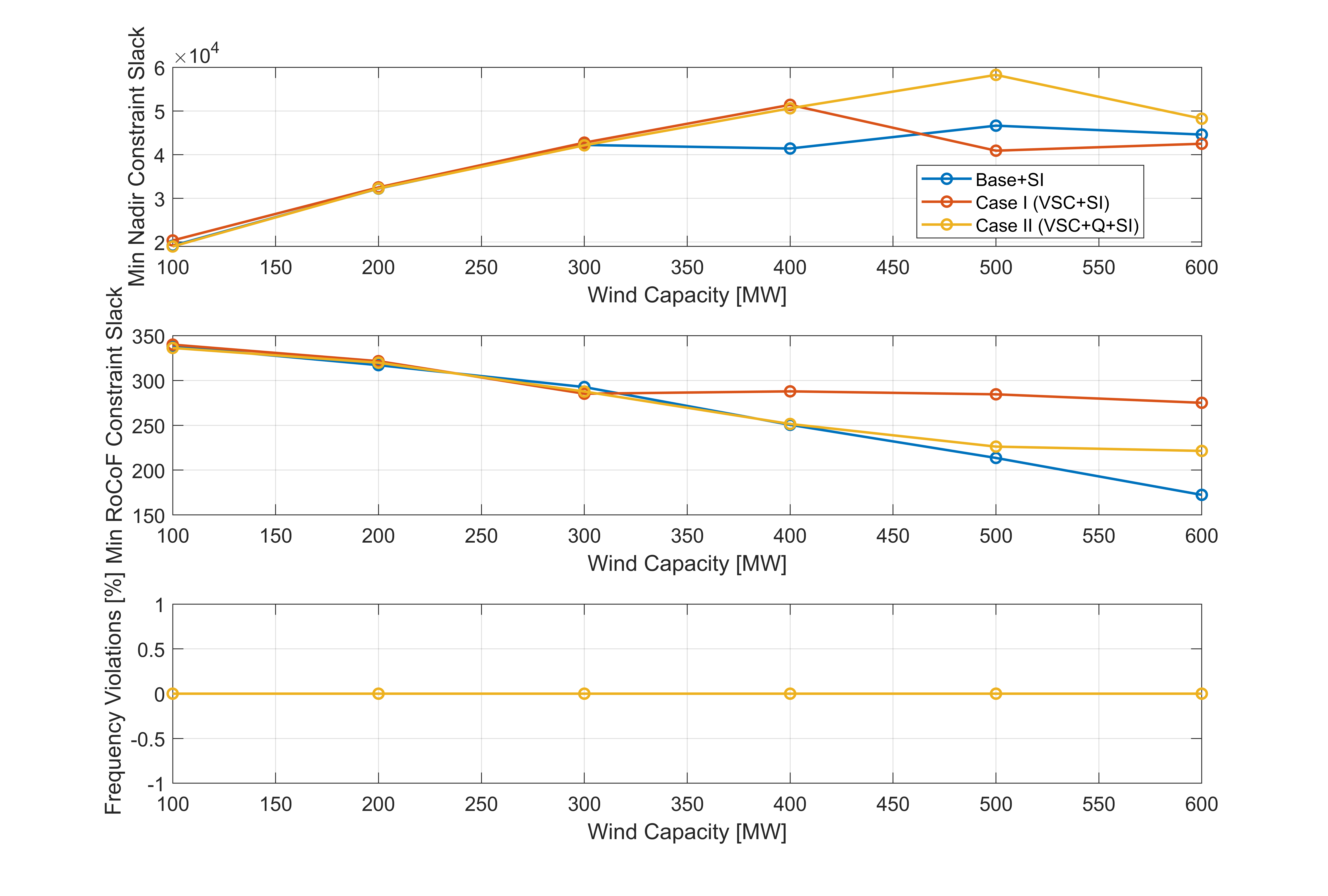}
    \vspace{-3mm}
    \caption{Minimum frequency-nadir slack, RoCoF slack, and violation rate across wind levels.}
    \label{fig:freq_metrics}
\end{figure}

\begin{figure}[ht]
    \centering
    \includegraphics[width=0.5\linewidth]{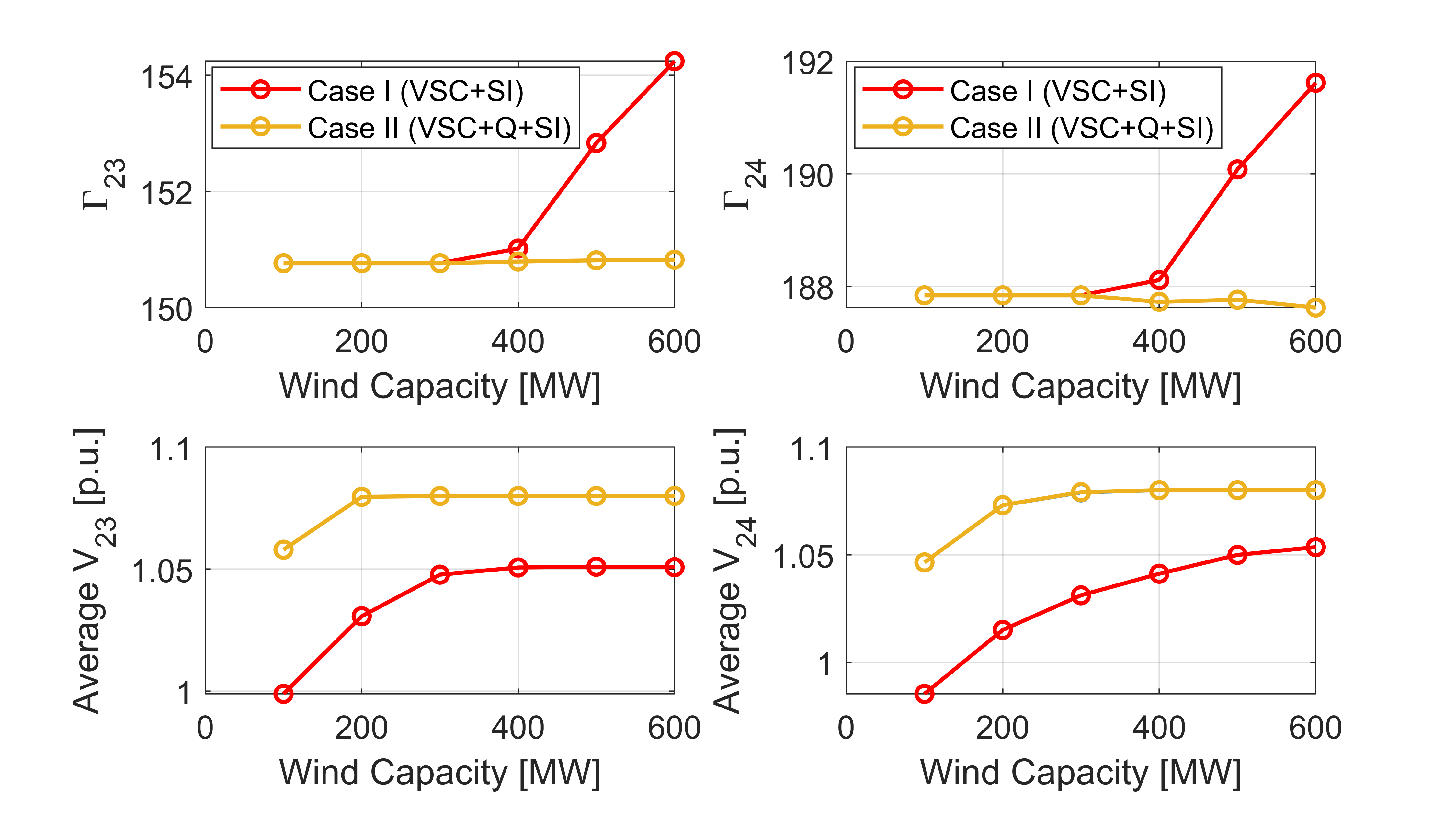}
    \vspace{-3mm}
    \caption{Voltage magnitude and stability margin at buses 23 and 24 versus wind penetration.}
    \label{fig:voltage_gamma}
\end{figure}

To further explain the operational differences among the three cases, additional results on curtailment, voltage support, and frequency performance are presented in Fig.~\ref{fig:curtail_shed}, Fig.~\ref{fig:freq_metrics}, and Fig.~\ref{fig:voltage_gamma}. Fig.~\ref{fig:curtail_shed} shows the expected wind curtailment and load shedding across the wind-penetration range. The Base+SI case exhibits negligible curtailment at all wind levels, since no voltage-stability constraint is enforced. In contrast, Case~I (VSC+SI) begins to curtail wind significantly beyond 400~MW, reaching about 100~MW at 600~MW wind. This reflects the need to maintain feasibility under the voltage-stability constraint without access to reactive-power support. Case~II (VSC+Q+SI) largely avoids this effect, with curtailment remaining below 30~MW even at high wind levels. In all cases, load shedding remains negligible (on the order of $10^{-10}$~MW), indicating that feasibility is maintained without relying on shedding.

Fig.~\ref{fig:freq_metrics} reports the minimum nadir slack, RoCoF slack, and frequency-violation rate. All three cases maintain zero frequency violations across the full wind range. The nadir and RoCoF slacks remain positive, confirming that the frequency-security constraints are consistently satisfied. This demonstrates that synthetic inertia is effectively embedded in the scheduling process, and that the differences observed earlier are not caused by a loss of frequency security but by voltage-stability limitations.

Fig.~\ref{fig:voltage_gamma} provides additional insight into voltage support at the critical IBG buses. In Case~I (VSC+SI), the voltage magnitudes at buses 23 and 24 increase with wind penetration as the optimization adjusts the operating point to satisfy the voltage-stability constraint. In Case~II (VSC+Q+SI), the voltages remain close to their upper limits across the full range due to the availability of reactive-power support from the IBGs. The corresponding stability margins $\Gamma_{23}$ and $\Gamma_{24}$ follow a similar trend, with Case~I (VSC+SI) requiring larger margins to maintain feasibility, while Case~II (VSC+Q+SI) achieves feasibility with smaller adjustments by using reactive support.

Overall, these results show that enforcing voltage stability without reactive flexibility leads to increased curtailment and more conservative operating points, while co-optimizing reactive power allows the system to maintain both security and economic efficiency.

\subsection{Representative Comparison at 400~MW Wind}

To provide a direct comparison, Table~\ref{tab:400mw_summary} summarizes the main operating metrics for the three cases at 400~MW installed wind capacity. This wind level is representative because it marks the point where the Base+SI case begins to show a substantial voltage-stability violation rate, while the constrained cases remain feasible.

The comparison highlights the trade-off between economy and security. Base+SI achieves the lowest operating cost at 6.395~k\$/h, but it also exhibits a voltage-stability violation rate of 45.83\%. Case~I (VSC+SI) eliminates these violations completely, but its cost increases to 7.233~k\$/h and the average wind curtailment rises to 21.35~MW. Case~II (VSC+Q+SI) also maintains zero voltage-stability violations, while reducing the operating cost to 6.511~k\$/h and keeping curtailment close to zero. This confirms that reactive-power support restores much of the flexibility lost in Case~I (VSC+SI).

The table also shows that frequency security is preserved in all three cases, with zero frequency violations at this operating point. In addition, the voltages at buses 23 and 24 remain stronger in Case~II (VSC+Q+SI) than in Case~I (VSC+SI), which is consistent with the improved local support provided by reactive-power control. Overall, the 400~MW case provides a compact illustration of the main result of the paper as voltage-stability constraints are necessary at higher renewable penetration, and reactive support significantly reduces their economic penalty.

\begin{table}[t]
\renewcommand{\arraystretch}{0.8}
\centering
\caption{Comparison of the three cases at 400~MW wind capacity.}
\vspace{-2mm}
\label{tab:400mw_summary}
\setlength{\tabcolsep}{3pt}
\resizebox{\columnwidth}{!}{
\begin{tabular}{||c|c|c|c|c|c|c|c|c||}
\hline \hline
\textbf{Case} & \textbf{Cost} & \textbf{Viol.} & \textbf{Curt.} & \textbf{Freq} & \textbf{$V_{23}$} & \textbf{$V_{24}$} & \textbf{$\Gamma_{23}$} & \textbf{$\Gamma_{24}$} \\
 & (k\$/h) & (\%) & (MW) & (\%) & (p.u.) & (p.u.) &  &  \\ \hline \hline
Base+SI            & 6.40 & 45.83 & 0.00  & 0.00 & 1.08 & 1.08 & N/A     & N/A     \\ \hline
Case~I (VSC+SI)             & 7.23 & 0.00  & 21.35 & 0.00 & 1.05 & 1.04 & 151.0 & 188.1 \\ \hline
Case~II (VSC+Q+SI)            & 6.51 & 0.00  & 0.00  & 0.00 & 1.08 & 1.08 & 150.8 & 187.7 \\ \hline \hline
\end{tabular}
}
\vspace{-4mm}
\end{table}

\subsection{Impact of STATCOM on Voltage Support}

\begin{figure}[ht]
    \centering
    \includegraphics[width=0.5\linewidth]{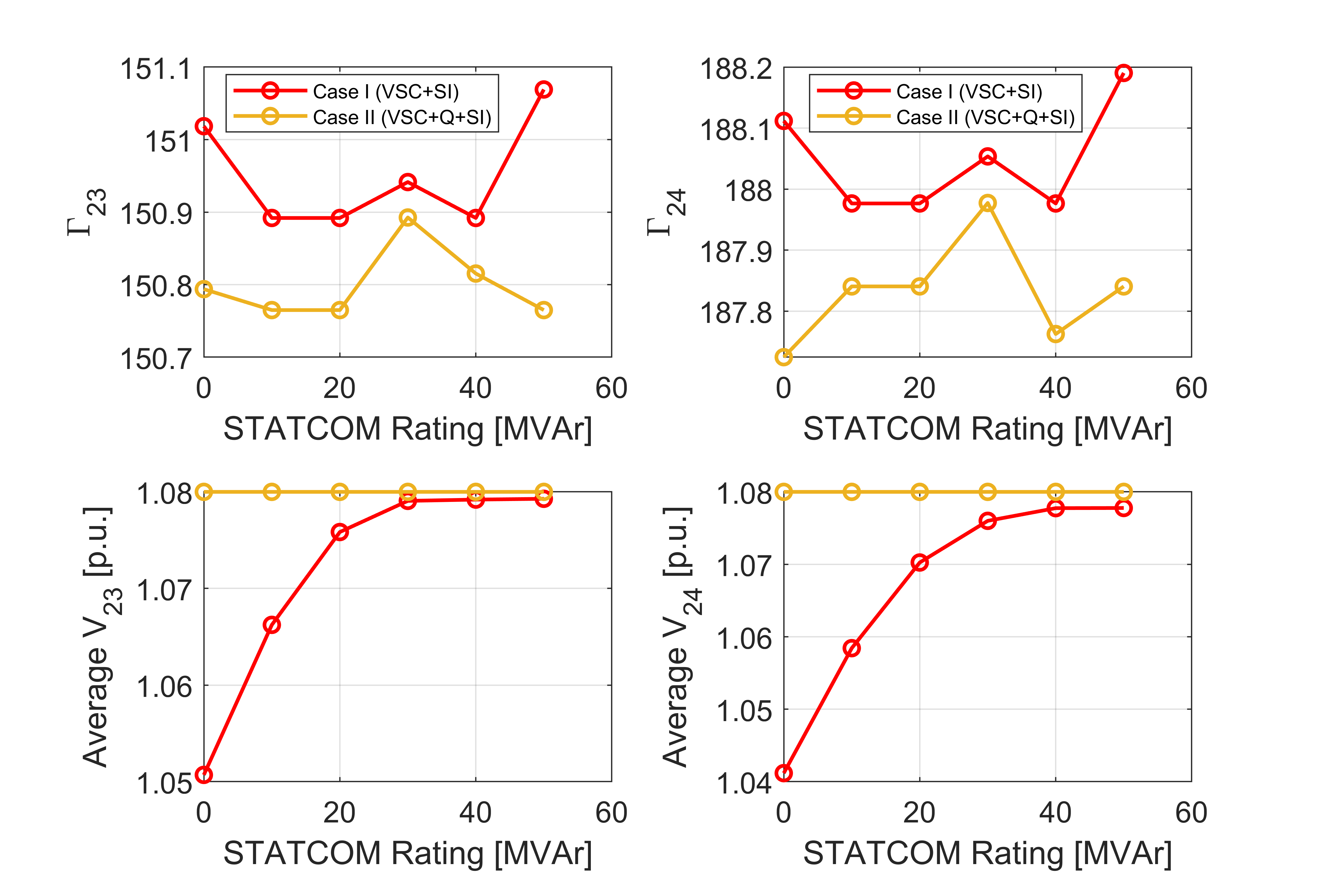}
    \vspace{-3mm}
    \caption{Effect of STATCOM rating on voltage-stability margins and bus voltages at buses 23 and 24 for the 400~MW wind case.}
    \label{fig:statcom_gamma_voltage}
\end{figure}

\begin{figure}[ht]
    \centering
    \includegraphics[width=0.5\linewidth]{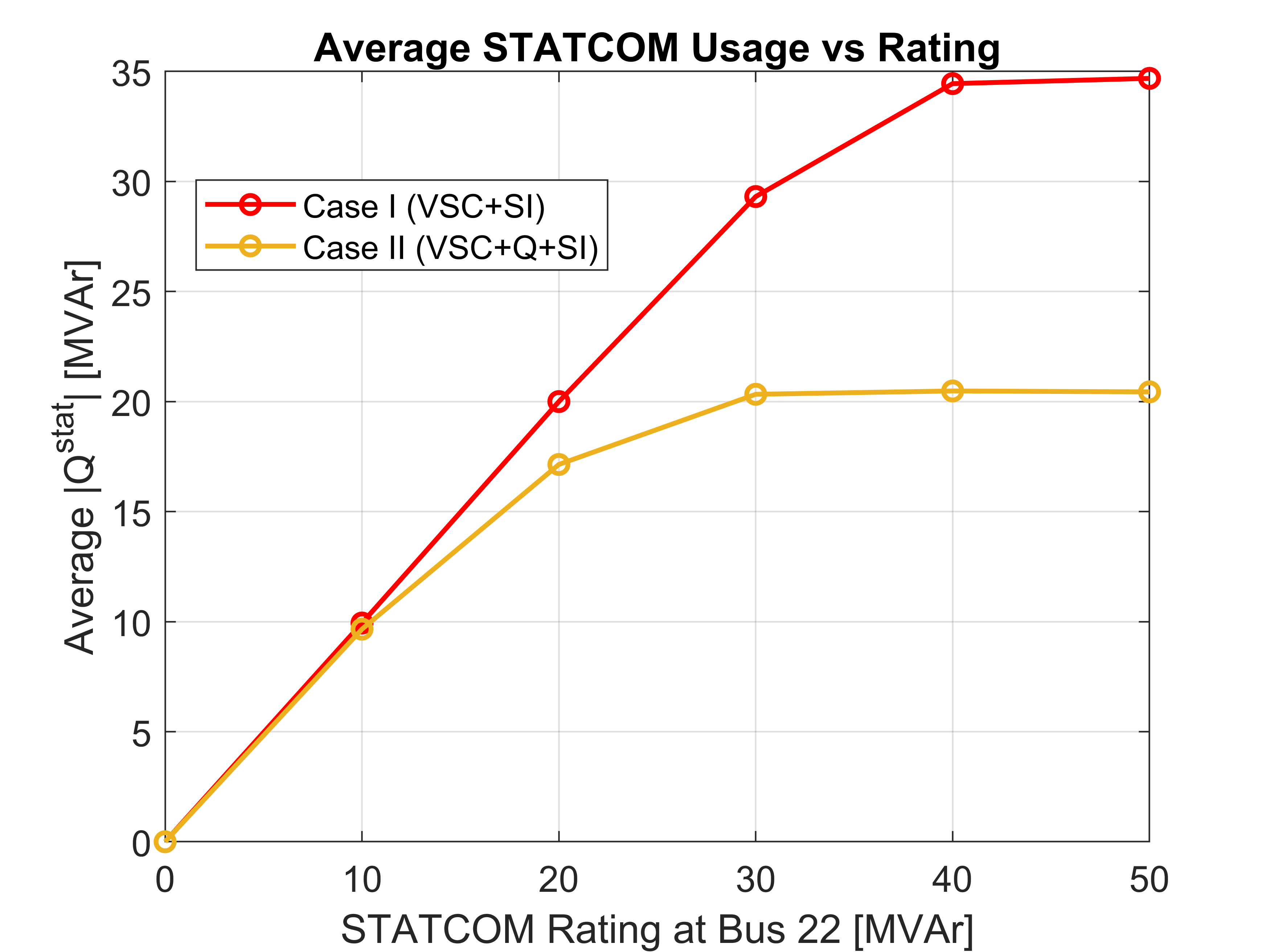}
    \vspace{-3mm}
    \caption{Average absolute STATCOM usage versus STATCOM rating for the 400~MW wind case.}
    \label{fig:statcom_usage}
\end{figure}

The operating cost trend in Fig.~\ref{fig:statcom_cost_bus22} is explained more clearly by the voltage-support results in Fig.~\ref{fig:statcom_gamma_voltage} and the STATCOM usage profile in Fig.~\ref{fig:statcom_usage}. These results are reported for the 400~MW wind case, which is representative because the unconstrained Base+SI schedule already exhibits a substantial voltage-stability violation rate at this operating point. Fig.~\ref{fig:statcom_gamma_voltage} shows the variation of the voltage-stability margins and bus voltages at the two critical IBG buses as the STATCOM rating increases. In Case~I (VSC+SI), increasing the STATCOM rating produces a clear improvement in the voltage profile at buses 23 and 24. The average voltages rise from about 1.05~p.u. and 1.04~p.u. without the STATCOM to values close to 1.08~p.u. once the rating reaches 30~MVAr. This indicates that the added reactive support directly relieves the stressed local operating condition created by the voltage-stability constraint. In Case~II (VSC+Q+SI), the voltages are already close to 1.08~p.u. even at low STATCOM ratings because IBG reactive-power support is available. As a result, increasing the STATCOM rating produces only a limited additional voltage benefit.

The corresponding $\Gamma_{23}$ and $\Gamma_{24}$ values show that the constrained schedules remain feasible throughout the rating range, but they do not change as strongly as the voltages. This suggests that, in the present model, the main benefit of the STATCOM is not a large increase in the learned stability margin itself, but a more favorable local voltage and reactive-power operating point. In other words, the STATCOM improves voltage support primarily through its direct reactive injection at the stressed area of the network.

Fig.~\ref{fig:statcom_usage} further explains why the cost curve in Fig.~\ref{fig:statcom_cost_bus22} flattens around 30~MVAr. In Case~I (VSC+SI), the average absolute STATCOM usage increases almost linearly up to 30~MVAr and then begins to saturate, indicating that most of the practically useful reactive support has already been captured by that rating. In Case~II (VSC+Q+SI), the average usage saturates even earlier, at around 20~MVAr, because part of the voltage-support requirement is already being supplied by the IBGs themselves. This behavior explains why larger STATCOM ratings provide only marginal additional operating-cost benefit. Taken together, these results support the choice of a 30~MVAr STATCOM as a practical reinforcement level. At this rating, the device captures most of the available voltage-support benefit in Case~I (VSC+SI), while larger capacities produce only limited additional improvement.

\subsection{STATCOM Siting Rationale}
\begin{figure}[ht]
    \centering
    \includegraphics[width=0.5\linewidth]{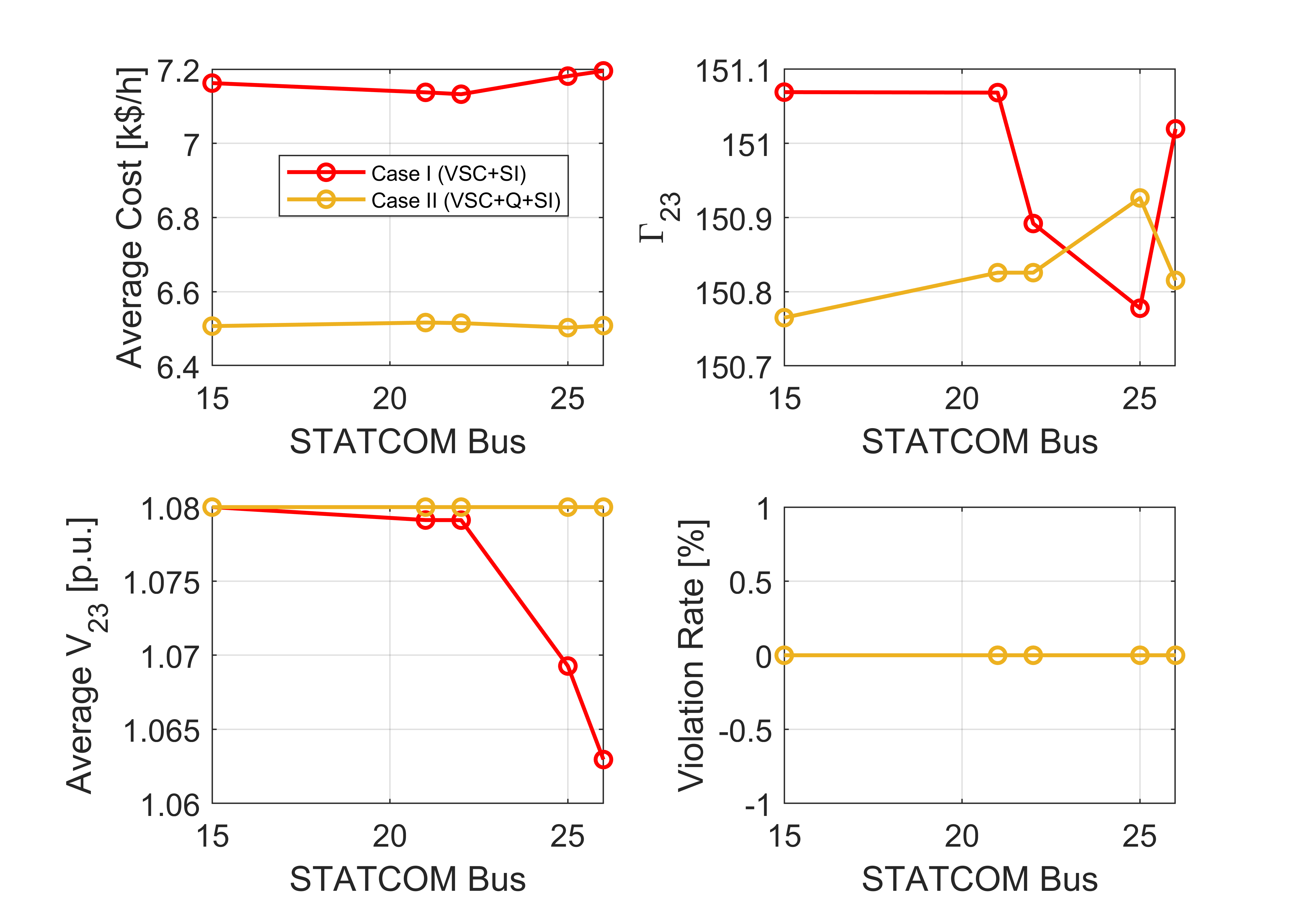}
    \vspace{-3mm}
    \caption{Effect of STATCOM siting on operating cost, voltage support, and violation rate for a 30~MVAr STATCOM at 400~MW wind.}
    \label{fig:statcom_siting}
\end{figure}

To verify that the choice of Bus~22 is not arbitrary, a siting sensitivity study is carried out for a fixed 30~MVAr STATCOM under the 400~MW wind case. Fig.~\ref{fig:statcom_siting} compares the effect of placing the STATCOM at several nearby buses on operating cost, voltage support, and voltage-stability performance.

The results show that the siting choice matters most in Case~I (VSC+SI), where the voltage-stability constraint is enforced without IBG reactive-power support. Among the tested buses, Bus~22 gives the lowest operating cost and one of the strongest voltage profiles at the critical IBG buses. Nearby buses such as 21 also perform well, which is consistent with their electrical proximity to the stressed area of the network. In contrast, placements farther from the weak IBG buses provide a smaller benefit, since the injected reactive power is less effective in supporting the local voltage condition that drives the stability constraint. In Case~II (VSC+Q+SI), the dependence on siting is weaker because part of the reactive-power support is already provided by the IBGs. As a result, the voltage profile remains strong and the operating cost changes only slightly across the tested locations. Even in this case, however, Bus~22 remains a competitive choice.

These results support the use of Bus~22 as a practical STATCOM location in the 30-bus study. It is electrically close to the weak IBG buses, gives one of the best operating-cost outcomes among the tested candidates, and provides effective local voltage support without the need for a larger device rating.

\subsection{Relaxation Quality and Computational Performance}

\begin{figure}[ht]
    \centering
    \includegraphics[width=0.5\linewidth]{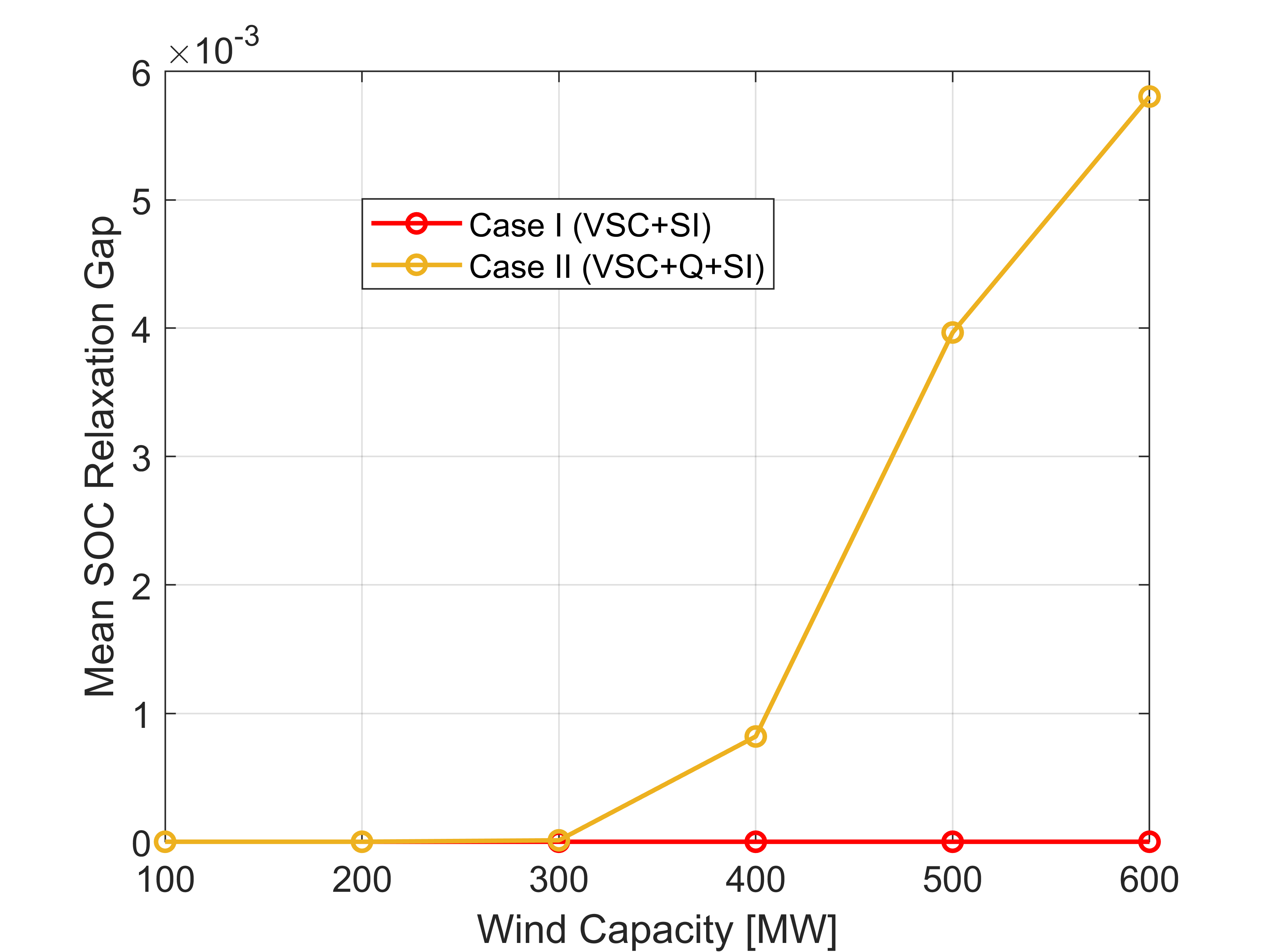}
    \vspace{-3mm}
    \caption{Mean SOC relaxation gap versus wind penetration for the VSC cases.}
    \label{fig:soc_gap}
\end{figure}

This subsection examines the tightness of the SOC relaxation and the computational burden of the proposed MISOCP formulation. All simulations are solved using the FICO Xpress optimizer with a relative optimality tolerance of 2\%, which provides a practical balance between solution quality and computational effort. Fig.~\ref{fig:soc_gap} shows the mean SOC relaxation gap as the installed wind capacity increases. The gap remains essentially zero in Case~I (VSC+SI) across the full wind range, which indicates a very tight relaxation for the voltage-stability-constrained formulation without IBG reactive-power support. In Case~II (VSC+Q+SI), the mean gap increases at higher wind penetration levels, especially beyond 400~MW, but it remains small in absolute terms, on the order of $10^{-3}$.
This indicates that the relaxation error remains small in magnitude for the studied operating conditions. However, since the network model is based on an SOC relaxation of the AC power flow equations, the reported gap should be interpreted as an indicator of relaxation tightness rather than as a direct guarantee of exact AC feasibility.

Table~\ref{tab:timing} reports the average solver time per rolling-horizon step for the three cases. As expected, Base+SI is the fastest because it does not include the voltage-stability cone constraints. Case~I (VSC+SI) requires additional SOC constraints associated with voltage stability, which increases the solution time. Case~II (VSC+Q+SI) is the most computationally demanding because it further includes IBG reactive-power decision variables and the associated coupling in the optimization model. Even with these additions, the reported solve times remain suitable for day-ahead scheduling studies.

Using a 2\% MIP gap does not change the main feasibility conclusions of the study. Instead, it reduces excessive computation in later branch-and-bound iterations while preserving the observed voltage-stability and frequency-security behavior of the schedules. Overall, these results indicate that the SOC relaxation remains tight for the present application, while the proposed formulation remains computationally tractable when solved with a commercial MISOCP solver.

\begin{table}[t]
\renewcommand{\arraystretch}{0.8}
\centering
\caption{Average solver time reported by Xpress per rolling-horizon step (one 24-hour SUC solve).}
\vspace{-3mm}
\label{tab:timing}
\begin{tabular}{||c|c||}
\hline \hline
\textbf{Case} & \textbf{Time (s/step)} \\ \hline \hline
Base+SI              & 21.693 \\ \hline
Case I (VSC+SI)      & 25.737 \\ \hline
Case II (VSC+Q+SI)   & 35.986 \\ \hline \hline
\end{tabular}
\vspace{-5mm}
\end{table}

\subsection{STATCOM Cost and TCO Assessment}
The objective in \eqref{eq:obj} includes only operating costs and load shedding. To provide economic context for the reinforcement at Bus~22, comparable benchmarks are summarized and a simple total-cost-of-ownership (TCO) model is evaluated outside the UC framework.

\noindent \textit{(1)} CAPEX: ISO-NE planning material lists indicate the vendor quotes of US\$13.0\,M for a $\pm$50~MVAr STATCOM and US\$14.5\,M for a $\pm$50~MVAr SC \cite{ISO-NE2021Stakeholder}, which would be even lower for a $\pm$30~MVAr STATCOM. The AER-approved South Australia system-strength project (four SCs) totals AU\$166\,M including civil works and auxiliaries \cite{AER2019SAnews}.\\
\textit{(2)} Losses: \cite{Neutz2013STATCOMPQ} shows STATCOM converter losses below 1\% at rated output, typical SC losses are in the 1.3-1.5\%, range depending on excitation and auxiliaries \cite{ABB2015SynconMining}.\\
\textit{(3)} O\&M convention: Where technology-specific data is unavailable, regulated planning studies use an annual O\&M proxy of 1\% of CAPEX \cite{AEMO2025RITTOandM}.

To estimate annual energy losses in \text{MWh}, the following loss and O\&M model is used in the TCO assessment:
Let \(S\) be the device rating (MVAr). Let \(\ell_{\text{stat}}\) and \(\ell_{\text{sc}}\) denote fractional losses at rated output. With baseMVA \(=100\)~MVA and \(8760\)~h/yr:\vspace{-3mm}
\begin{align}
P_{\text{loss}}^{\text{stat}} &= \ell_{\text{stat}}\, S \quad \text{(MW)}, \qquad
E_{\text{loss}}^{\text{stat}} = 8760\, P_{\text{loss}}^{\text{stat}} \quad \text{(MWh/yr)} \\
P_{\text{loss}}^{\text{sc}}   &= \ell_{\text{sc}}\, S \quad \text{(MW)}, \qquad
E_{\text{loss}}^{\text{sc}}   = 8760\ P_{\text{loss}}^{\text{sc}} \quad \text{(MWh/yr)}
\end{align}
For an energy price \(p_{\text{e}}\) [\$/MWh], the annual loss cost and annual O\&M are:\vspace{-3mm}
\begin{align}
    C_{\text{loss}} = p_{\text{e}}\,E_{\text{loss}}, \qquad
    C_{\text{O\&M}} = 0.01\,\text{CAPEX}
\end{align}

These terms are not part of the UC objective function, and can be evaluated separately for TCO assessment. Using \(\ell_{\text{stat}}=0.8\%\) and \(\ell_{\text{sc}}=1.5\%\):\vspace{-3mm}
\begin{align}
P_{\text{loss}}^{\text{stat}} &= 0.008\times 30 = 0.24~\text{MW} \\
E_{\text{loss}}^{\text{stat}} &= 8760\times 0.24 = 2102~\text{MWh/yr} \\
P_{\text{loss}}^{\text{sc}}   &= 0.015\times 30 = 0.45~\text{MW} \\
E_{\text{loss}}^{\text{sc}}   &= 8760\times 0.45 = 3942~\text{MWh/yr}
\end{align}
These results support the selection of a 30~MVAr STATCOM at Bus~22, as it captures the operating-cost benefits observed in Fig.~\ref{fig:statcom_cost_bus22} without increasing short-circuit levels in the adopted model.

\subsection{Scalability Analysis on IEEE 118-Bus System}
To examine whether the proposed formulation remains meaningful on a larger network, an additional study is carried out on the modified IEEE 118-bus system. The same modeling framework is retained, including the impedance-based voltage-stability constraints, the SOC network formulation, and the three operating cases considered earlier. A 24-hour simulation is used to compare the commitment and dispatch behavior across the larger system.
In this study, inverter-based resources (IBRs) are introduced to create operating conditions with high renewable penetration. Specifically, wind generation is integrated at buses 3, 41, 72, and 87, with a total installed capacity of 4000~MW distributed across these locations. These buses are selected to represent geographically dispersed injection points within the network, ensuring that the impact of renewable integration on system strength and voltage stability can be effectively captured. The overall penetration level is chosen to be sufficiently high so that voltage-stability constraints become active during system operation.

\begin{figure}[ht]
    \centering
    \includegraphics[width=0.6\linewidth]{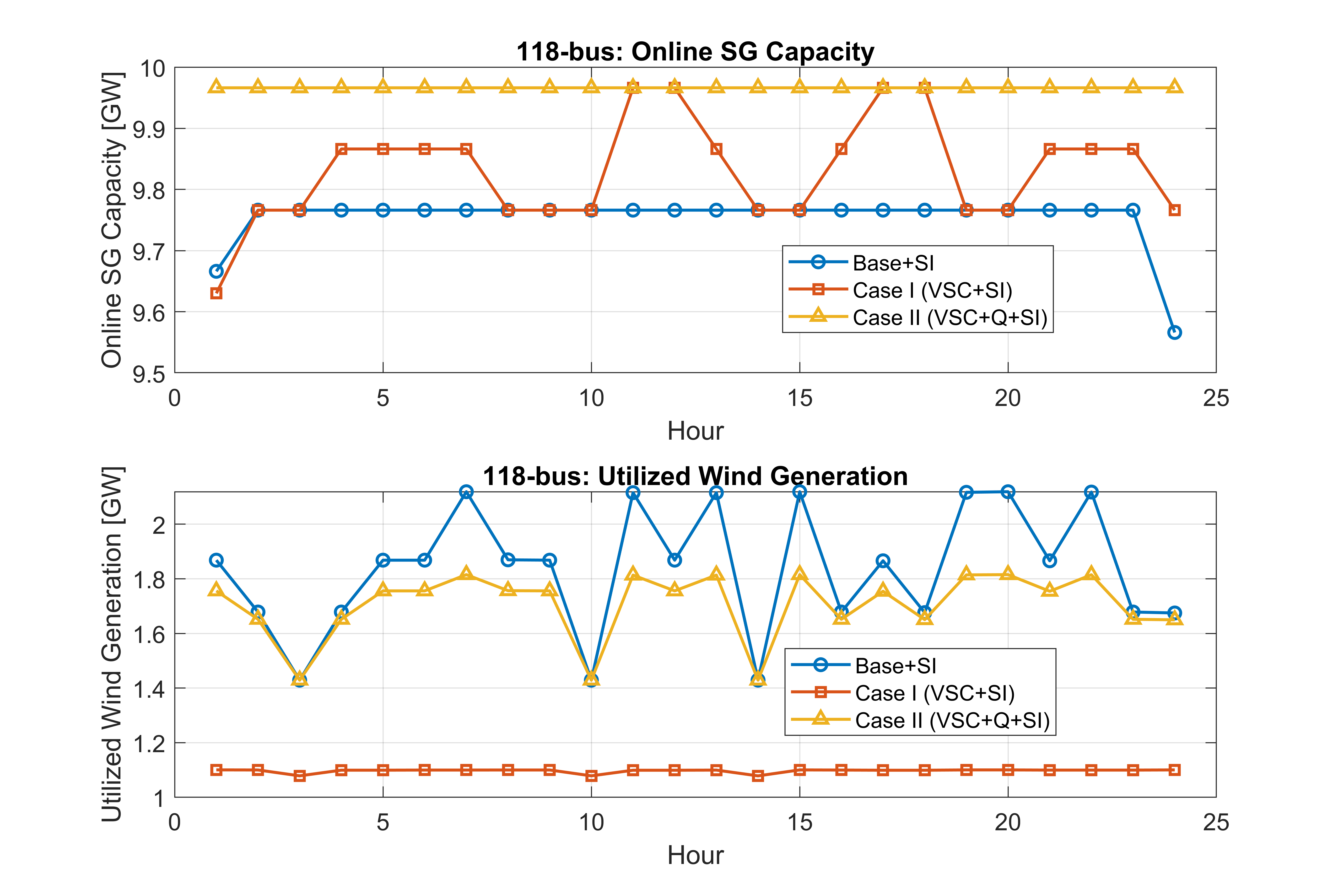}
    \vspace{-3mm}
    \caption{Hourly online synchronous-generation capacity and utilized wind generation in the modified IEEE 118-bus system over a 24-hour horizon.}
    \label{fig:118bus_results}
\end{figure}

Fig.~\ref{fig:118bus_results} shows the hourly online synchronous-generation capacity and the utilized wind generation over the 24-hour horizon. The upper plot indicates that Case~II (VSC+Q+SI) keeps the largest amount of synchronous capacity online throughout the day, followed by Case~I (VSC+SI), while the Base+SI case commits the smallest synchronous capacity on average. This is consistent with the role of the voltage-stability constraints. Once the VSC constraints are enforced, the schedule becomes more conservative and commits more synchronous capacity to support system strength. The effect is strongest in Case~II (VSC+Q+SI), where both voltage-stability enforcement and reactive-power flexibility shape the operating point.

The lower plot gives the clear view of how the larger system behaves under the three formulations. As expected, the Base+SI case utilizes the largest amount of wind generation because no voltage-stability constraint is imposed. Case~I (VSC+SI) has the lowest wind utilization across the full horizon, since the voltage-stability constraint must be satisfied without IBG reactive-power support. Case~II (VSC+Q+SI) remains between the other two cases. It still respects the voltage-stability constraint, but reactive-power support allows it to recover a substantial portion of the renewable utilization lost in Case~I (VSC+SI). This is the same qualitative pattern observed in the IEEE 30-bus study. In other words, the larger system confirms the same main conclusion that enforcing voltage stability alone reduces renewable utilization, while adding reactive-power flexibility helps recover part of that loss.

Table~\ref{tab:118_scalability} summarizes the expected operating cost, wind curtailment, voltage-stability violation rate, and solver time for the three cases. The Base+SI case gives the lowest expected cost, but this comes at the expense of a high voltage-stability violation rate of 79.17\%, indicating frequent infeasible operating conditions despite relatively low curtailment (249.16~MW on average). 
Case~I (VSC+SI) fully eliminates voltage-stability violations, but at the cost of significantly increased renewable curtailment, reaching 989.87~MW on average. This reflects the conservative dispatch required when voltage stability is enforced without IBG reactive-power support. 
Case~II (VSC+Q+SI) also achieves zero violation rate while substantially reducing curtailment compared to Case~I (VSC+SI), with an average of 381.46~MW. This demonstrates that reactive-power flexibility allows a better balance between maintaining voltage stability and utilizing available renewable generation. 
The solver times follow the expected trend: the Base+SI case is the fastest, while Case~I (VSC+SI) and especially Case~II (VSC+Q+SI) require more computation due to the additional voltage-stability and reactive-power decision variables in the larger IEEE 118-bus system. Overall, the 118-bus results show that the proposed formulation is not restricted to the smaller benchmark. The same operational trade-offs observed in the 30-bus system remain visible at the larger scale, and the model can still be solved using the same MISOCP framework.

It is also worth noting that the formulation can accommodate multiple inverter-based resources without structural changes. Increasing the number of IBGs mainly affects the number of monitored voltage-stability constraints, while the underlying model structure remains unchanged. Therefore, similar qualitative behavior is expected, although the computational burden increases as the number of monitored buses grows.

However, it should be emphasized that the present 118-bus study primarily demonstrates scalability with respect to network size rather than providing a systematic sensitivity analysis with respect to the number of IBRs. In particular, the number and placement of IBGs are fixed in this setup, and no parametric variation of IBR count is performed. A dedicated study that explicitly varies the number of IBRs while keeping other system characteristics controlled is left for future work.

\begin{table}[t]
\renewcommand{\arraystretch}{0.8}
\centering
\caption{Computational performance and system-level outcomes for the IEEE 118-bus scalability study.}
\vspace{-3mm}
\label{tab:118_scalability}
\setlength{\tabcolsep}{4pt}
\resizebox{\columnwidth}{!}{
\begin{tabular}{||c|c|c|c|c|c|c||}
\hline \hline
\textbf{Case} & \textbf{Cost} & \textbf{Avail.} & \textbf{Used} & \textbf{Curt.} & \textbf{VS Viol.} & \textbf{Time} \\
              & \textbf{(k\$/h)} & \textbf{(MW)} & \textbf{(MW)} & \textbf{(MW)} & \textbf{(\%)} & \textbf{(s)} \\ \hline \hline
Base+SI    & 1613.65 & 2087.31 & 1838.15 & 249.16 & 79.17 & 29.55 \\ \hline
Case~I (VSC+SI)  & 2109.68 & 2087.31 & 1097.44 & 989.87 & 0.00  & 84.31 \\ \hline
Case~II (VSC+Q+SI) & 1762.65 & 2087.31 & 1705.85 & 381.46 & 0.00  & 117.13 \\ \hline \hline
\end{tabular}
}
\vspace{-4mm}
\end{table}

\subsection{Comparison with Synchronous Condenser Reinforcement}
\label{sec:sc_comparison}

To provide a direct comparison with the SC-based reinforcement framework in~\cite{chu2023voltage}, an additional study is carried out by replacing the proposed STATCOM reinforcement with a synchronous condenser (SC) placed at Bus~22. The SC is introduced at the same electrically relevant location used in the STATCOM study, and its capacity is varied as $\bar Q_{\mathrm{sc}}\in\{0,10,20,30,40,50\}$~MVAr under the same 400~MW wind condition. 
The SC is modeled with both its reactive-power capability and its admittance contribution to system strength, consistent with~\cite{chu2023voltage}, while its inertia is not explicitly modeled because the frequency-nadir constraint is already satisfied by synthetic inertia in all reported cases.
The purpose of this comparison is not to change the role of the voltage-stability constraints, but to examine how an SC affects the operating point when the same VSC-based scheduling framework is used. As in the rest of the paper, zero voltage-stability violations in Case~I (VSC+SI) and Case~II (VSC+Q+SI) are achieved because the voltage-stability cone is enforced in the optimization. The SC does not replace the role of the voltage-stability constraints. It only modifies the operating point of the system while those constraints are enforced.

\begin{figure}[ht]
    \centering
    \includegraphics[width=0.5\linewidth]{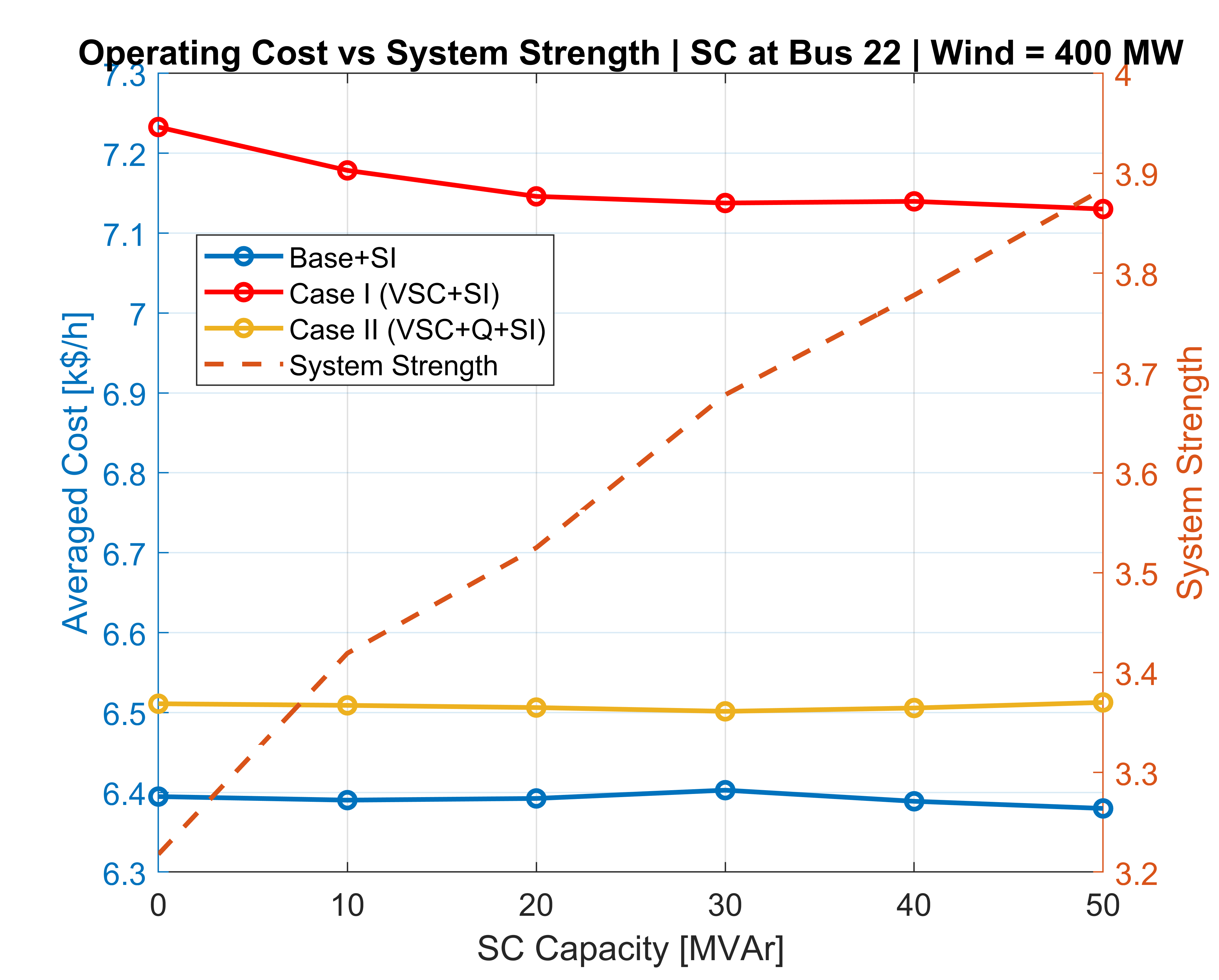}
    \vspace{-3mm}
    \caption{Impact of SC capacity at Bus~22 on average operating cost at 400~MW wind. The dashed curve on the right axis shows the corresponding increase in average system strength.}
    \label{fig:sc_cost_bus22}
\end{figure}

Fig.~\ref{fig:sc_cost_bus22} shows the variation of average operating cost with SC capacity for the three cases. The Base+SI case remains nearly unchanged and continues to exhibit a high voltage-stability violation rate, since the voltage-stability constraint is not enforced in that case. In contrast, the SC has a visible effect in the two VSC-constrained cases, especially in Case~I (VSC+SI). When no SC is installed, Case~I (VSC+SI) has the highest operating cost because the schedule must satisfy the voltage-stability constraint without support from IBG reactive power. As the SC capacity increases, the cost of Case~I (VSC+SI) decreases gradually and then becomes nearly flat beyond about 30~MVAr. This indicates that the main operating benefit of SC reinforcement is already captured around this level, while further increases in capacity provide only limited additional economic improvement. Case~II (VSC+Q+SI) shows a much smaller sensitivity to the SC size because reactive-power support is already available from the IBGs. These results are consistent with the earlier STATCOM study, where the largest reinforcement benefit also appeared in Case~I (VSC+SI) and saturated around the same practical rating.

\begin{figure}[ht]
    \centering
    \includegraphics[width=0.5\linewidth]{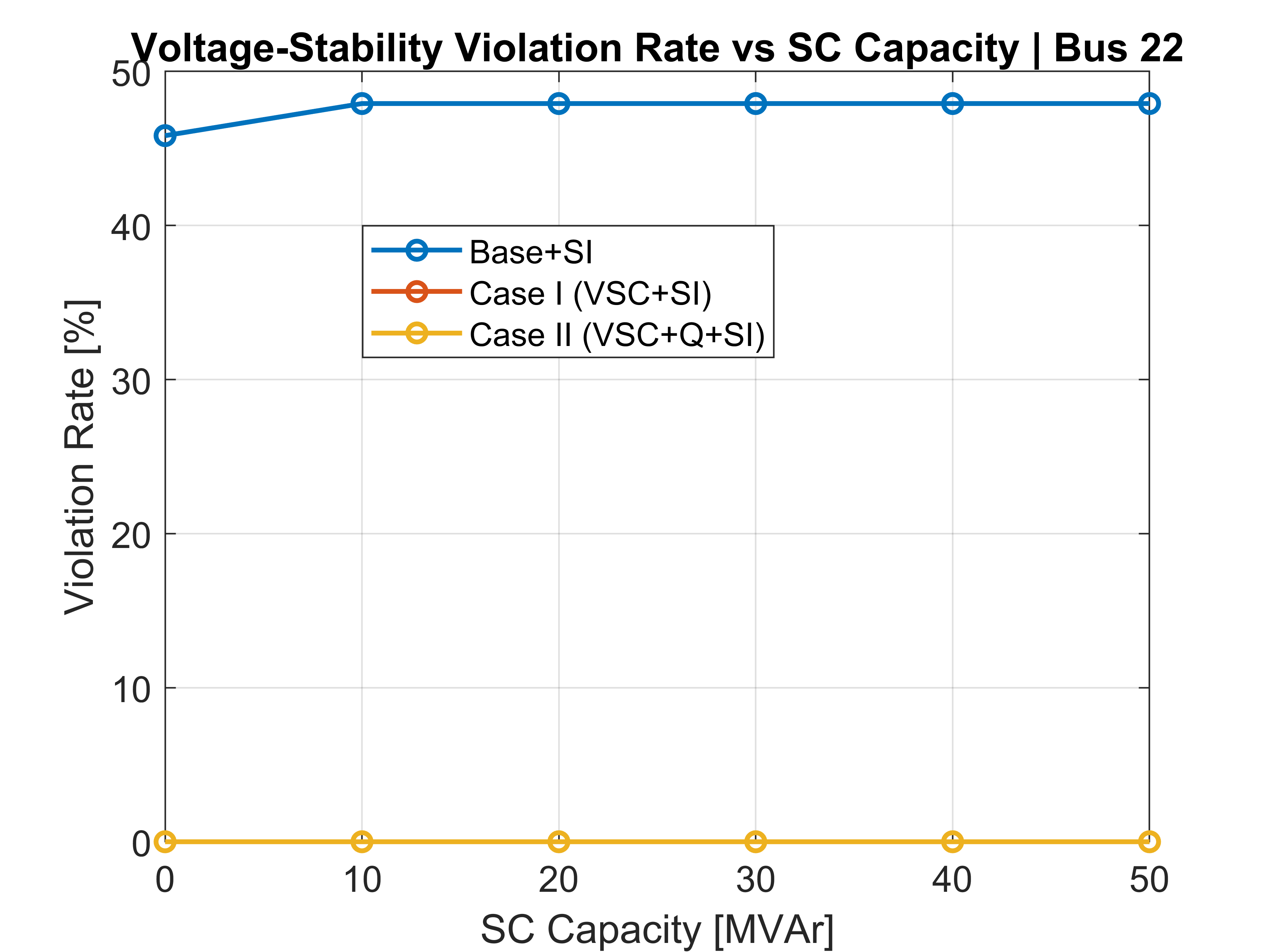}
    \vspace{-3mm}
    \caption{Voltage-stability violation rate versus SC capacity at Bus~22 for the 400~MW wind case.}
    \label{fig:sc_violation_bus22}
\end{figure}

The voltage-stability violation-rate results are reported in Fig.~\ref{fig:sc_violation_bus22}. The figure confirms the same fundamental point observed throughout the paper. The Base+SI case remains highly insecure, with a violation rate close to 46--48\% across the full SC range, because the voltage-stability cone is not part of the optimization in that case. By contrast, both Case~I (VSC+SI) and Case~II (VSC+Q+SI) remain at zero violation rate for all SC capacities because the VSC formulation is enforced directly. Therefore, the benefit of the SC should not be interpreted as eliminating violations on its own. Its role is instead to improve the operating condition of the constrained schedule, which can then reduce the economic burden of maintaining voltage stability.

\begin{figure}[ht]
    \centering
    \includegraphics[width=0.6\linewidth]{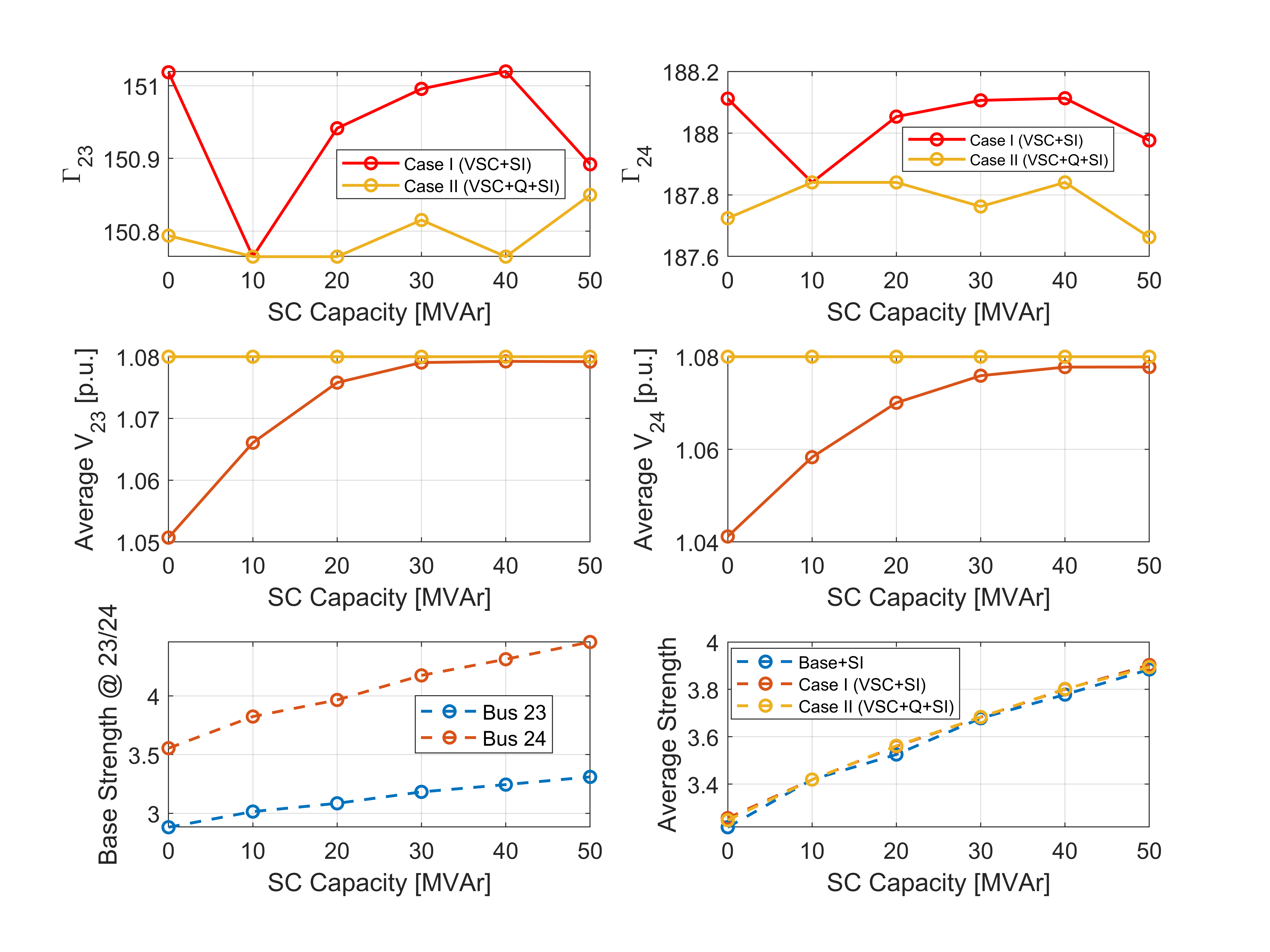}
    \vspace{-3mm}
    \caption{Effect of SC capacity at Bus~22 on stability margins, bus voltages, and system-strength indicators for the 400~MW wind case.}
    \label{fig:sc_gamma_voltage_strength}
\end{figure}

The mechanism behind this behavior is illustrated more clearly in Fig.~\ref{fig:sc_gamma_voltage_strength}. The upper two plots show the average stability margins $\Gamma_{23}$ and $\Gamma_{24}$ in the two VSC-constrained cases. These quantities remain in a narrow feasible range as the SC rating changes, which indicates that all reported schedules remain inside the secure region. The middle two plots show the corresponding voltages at buses~23 and~24. In Case~I (VSC+SI), the average voltage magnitude at both buses increases clearly as the SC capacity grows, especially from 0 to 30~MVAr. This means that the SC relieves the stressed local operating condition and allows the schedule to maintain voltage stability with a less conservative dispatch. In Case~II (VSC+Q+SI), the voltages are already close to their upper operating range because reactive-power support from the IBGs is available, so the additional benefit of the SC is smaller.

The lower plots of Fig.~\ref{fig:sc_gamma_voltage_strength} provide the main physical distinction between the SC and the STATCOM reinforcement. Unlike the STATCOM model used in this paper, the SC increases system strength as its capacity grows. This can be seen both in the bus-level strength indicators and in the average system-strength measure, which increase almost monotonically with the SC size. Therefore, the SC supports the weak operating point not only through reactive-power capability, but also by increasing short-circuit strength. This is useful to show because it clarifies why the SC can improve the constrained operating point, and at the same time it highlights the main practical difference from the STATCOM-based reinforcement. The STATCOM achieved its benefit mainly through local reactive support, whereas the SC changes the strength level of the surrounding network.

To provide a representative comparison, Table~\ref{tab:sc30_summary} summarizes the main metrics at 30~MVAr SC capacity, which is the same practical rating selected earlier for the STATCOM discussion. This rating is sufficient to capture most of the operating-cost benefit while avoiding unnecessary oversizing. The table shows that, at this operating point, the Base+SI case still has the lowest operating cost but remains highly voltage-insecure. Case~I (VSC+SI) and Case~II (VSC+Q+SI) both maintain zero voltage-stability violations, as expected from the enforced VSC model. The SC reduces the operating burden in Case~I (VSC+SI) by improving the voltage profile and supporting a stronger operating point at the weak buses, which is consistent with the cost reduction trend observed in Fig.~\ref{fig:sc_cost_bus22}. Case~II (VSC+Q+SI) remains the most economical secure case because it combines enforced voltage stability with IBG reactive-power flexibility. The voltage and $\Gamma$ values in the table are consistent with the trends seen in Fig.~\ref{fig:sc_gamma_voltage_strength}, where the SC improves the local operating condition but does not materially change the main ordering of the three cases.

\begin{table}[t]
\renewcommand{\arraystretch}{0.85}
\centering
\caption{Comparison of the three cases with a 30~MVAr synchronous condenser at Bus~22 under 400~MW wind.}
\vspace{-2mm}
\label{tab:sc30_summary}
\setlength{\tabcolsep}{3.5pt}

\resizebox{\columnwidth}{!}{
\begin{tabular}{||c|c|c|c|c|c|c|c|c|c||}
\hline \hline
\textbf{Case} & \textbf{Cost} & \textbf{Viol.} & \textbf{Curt.} & \textbf{Freq} & \textbf{$V_{23}$} & \textbf{$V_{24}$} & \textbf{$\Gamma_{23}$} & \textbf{$\Gamma_{24}$} & \textbf{Strength} \\
 & (k\$/h) & (\%) & (MW) & (\%) & (p.u.) & (p.u.) &  &  &  \\ \hline \hline
Base+SI            & 6.403 & 47.92 & 0.00   & 0.00 & 1.0800 & 1.0800 & N/A     & N/A     & 3.678 \\ \hline
Case~I (VSC+SI)    & 7.138 & 0.00  & 21.216 & 0.00 & 1.0791 & 1.0759 & 150.996 & 188.106 & 3.683 \\ \hline
Case~II (VSC+Q+SI) & 6.502 & 0.00  & 0.00   & 0.00 & 1.0800 & 1.0800 & 150.815 & 187.762 & 3.683 \\ \hline \hline
\end{tabular}
}

\vspace{-4mm}
\end{table}

The solver-time results at 30~MVAr reported in Table~\ref{tab:sc_timing} remain consistent with the rest of the paper. The Base+SI case is the fastest because it does not include the voltage-stability cone, while Case~I (VSC+SI) and Case~II (VSC+Q+SI) require additional computational effort due to the VSC formulation and reactive-power coupling. Compared with the corresponding STATCOM results in Table~\ref{tab:timing}, the SC-based cases are noticeably slower, which is expected because the SC modifies $Y^g$ and therefore changes the $Z$-ratios used in the regression-based linearization, whereas the STATCOM enters only the reactive nodal balance without altering that part of the model. However, the ordering of the solve times remains the same, with Base+SI fastest, followed by Case~I (VSC+SI), and then Case~II (VSC+Q+SI).

Overall, the SC study provides a fair operational benchmark against the framework in~\cite{chu2023voltage}. The results show that an SC at Bus~22 improves the constrained operating point and reduces the operating penalty associated with voltage-stability enforcement, particularly in Case~I (VSC+SI). However, the comparison with the earlier STATCOM results highlights an important practical distinction. The SC achieves this benefit by increasing system strength and short-circuit contribution, whereas the STATCOM delivers a comparable operating benefit through local reactive support without increasing short-circuit levels in the adopted model. When this observation is combined with the TCO analysis in Section~5.9, which indicates a lower expected ownership burden and lower losses for the STATCOM under typical operating assumptions, the results support the use of the proposed STATCOM reinforcement as a practical and economically favorable extension of the SC-based framework.

\begin{table}[t]
\renewcommand{\arraystretch}{0.85}
\centering
\caption{Average solver time per rolling-horizon step for the SC reinforcement study.}
\vspace{-2mm}
\label{tab:sc_timing}
\begin{tabular}{||c|c||}
\hline \hline
\textbf{Case} & \textbf{Time (s/step)} \\ \hline \hline
Base+SI            & 21.316 \\ \hline
Case I (VSC+SI)    & 34.949 \\ \hline
Case II (VSC+Q+SI) & 50.899 \\ \hline \hline
\end{tabular}
\vspace{-4mm}
\end{table}

\vspace{-3mm}
\section{Conclusions}
\label{sec:6_Conclusions}
\vspace{-3mm}

This paper presents a VSC-UC framework for power systems with high penetration of inverter-based generation. The model embeds impedance-based voltage stability constraints in SOC form within a rolling-horizon MISOCP formulation, allowing voltage and frequency security to be enforced jointly under uncertainty. A key contribution is the integration of STATCOM-based reactive power support directly into the unit commitment problem. The STATCOM is modeled through symmetric reactive limits and a voltage-dependent constraint, enabling its impact to be assessed consistently alongside generator commitment, inverter dispatch, and synthetic inertia. The results show that voltage stability becomes a limiting factor at high renewable penetration and leads to increased cost and curtailment if not properly managed. Incorporating reactive power support reduces the operating penalty of voltage-stability enforcement and recovers a significant portion of renewable utilization. The analysis also identifies a practical STATCOM rating beyond which additional capacity provides limited benefit. A direct comparison with synchronous-condenser reinforcement shows that both devices can improve the constrained operating point, while the STATCOM retains the practical advantage of avoiding increased short-circuit levels in the adopted model. Finally, the SOC relaxation remains tight, and the formulation is computationally tractable and scalable, as demonstrated on the IEEE 118-bus system. Overall, the proposed framework provides a practical approach for co-optimizing commitment, frequency support, and voltage stability in modern power systems.

\vspace{-3mm}

\bibliographystyle{elsarticle-num} 
\bibliography{references}

@INPROCEEDINGS{milano2018foundations,
  author={Milano, Federico and Dörfler, Florian and Hug, Gabriela and Hill, David J. and Verbič, Gregor},
  booktitle={2018 Power Systems Computation Conference (PSCC)}, 
  title={Foundations and Challenges of Low-Inertia Systems (Invited Paper)}, 
  year={2018},
  volume={},
  number={},
  pages={1-25},
  doi={10.23919/PSCC.2018.8450880}}

@article{modarresi2016review,
title = {A comprehensive review of the voltage stability indices},
journal = {Renewable and Sustainable Energy Reviews},
volume = {63},
pages = {1-12},
year = {2016},
issn = {1364-0321},
doi = {https://doi.org/10.1016/j.rser.2016.05.010},
author = {Javad Modarresi and Eskandar Gholipour and Amin Khodabakhshian}
}

@ARTICLE{wu2018scr,
  author={Wu, Di and Li, Gangan and Javadi, Milad and Malyscheff, Alexander M. and Hong, Mingguo and Jiang, John Ning},
  journal={IEEE Transactions on Sustainable Energy}, 
  title={Assessing Impact of Renewable Energy Integration on System Strength Using Site-Dependent Short Circuit Ratio}, 
  year={2018},
  volume={9},
  number={3},
  pages={1072-1080},
  doi={10.1109/TSTE.2017.2764871}}

@ARTICLE{chu2021short,
  author={Chu, Zhongda and Teng, Fei},
  journal={IEEE Transactions on Power Systems}, 
  title={Short Circuit Current Constrained UC in High IBG-Penetrated Power Systems}, 
  year={2021},
  volume={36},
  number={4},
  pages={3776-3785},
  doi={10.1109/TPWRS.2021.3053074}}

@INPROCEEDINGS{chutengSCC,
  author={Chu, Zhongda and Teng, Fei},
  booktitle={2025 IEEE Kiel PowerTech}, 
  title={Impact of IBR Modeling on SCC Calculation and SCC-Constrained System Operation}, 
  year={2025},
  volume={},
  number={},
  pages={1-6},
  doi={10.1109/PowerTech59965.2025.11180470}}

@ARTICLE{chu2023voltage,
  author={Chu, Zhongda and Teng, Fei},
  journal={IEEE Transactions on Power Systems}, 
  title={Voltage Stability Constrained Unit Commitment in Power Systems With High Penetration of Inverter-Based Generators}, 
  year={2023},
  volume={38},
  number={2},
  pages={1572-1582},
  doi={10.1109/TPWRS.2022.3179563}}

@article{Alajrash2024EnergyReports,
title = {A comprehensive review of FACTS devices in modern power systems: Addressing power quality, optimal placement, and stability with renewable energy penetration},
journal = {Energy Reports},
volume = {11},
pages = {5350-5371},
year = {2024},
issn = {2352-4847},
doi = {https://doi.org/10.1016/j.egyr.2024.05.011},
author = {Ban H. Alajrash and Mohamed Salem and Mahmood Swadi and Tomonobu Senjyu and Mohamad Kamarol and Saad Motahhir}
}

@article{FACTS_SCUC,
title = {Analysis of FACTS devices on Security Constrained Unit Commitment problem},
journal = {International Journal of Electrical Power \& Energy Systems},
volume = {66},
pages = {280-293},
year = {2015},
issn = {0142-0615},
doi = {https://doi.org/10.1016/j.ijepes.2014.10.049},
author = {S. Sreejith and Sishaj P. Simon and M.P. Selvan}
}

@ARTICLE{freq-nadir-UC,
  author={Liu, Xuebo and Fang, Xin and Gao, Ningchao and Yuan, Haoyu and Hoke, Andy and Wu, Hongyu and Tan, Jin},
  journal={IEEE Open Access Journal of Power and Energy}, 
  title={Frequency Nadir Constrained Unit Commitment for High Renewable Penetration Island Power Systems}, 
  year={2024},
  volume={11},
  number={},
  pages={141-153},
  doi={10.1109/OAJPE.2024.3370504}}

@misc{ISO-NE2021Stakeholder,
  title        = {Dynamic Reactive Device Technologies Stakeholder Feedback},
  author       = {{ISO New England}},
  year         = {2021},
  howpublished = {\url{https://www.iso-ne.com/static-assets/documents/2021/04/a6_dynamic_reactive_device_technologies_stakeholder_feedback.pdf}},
}

@article{kocuk2016strong,
   title={Strong SOCP Relaxations for the Optimal Power Flow Problem},
   volume={64},
   ISSN={1526-5463},
   DOI={10.1287/opre.2016.1489},
   number={6},
   journal={Operations Research},
   publisher={Institute for Operations Research and the Management Sciences (INFORMS)},
   author={Kocuk, Burak and Dey, Santanu S. and Sun, X. Andy},
   year={2016},
   month=dec, pages={1177–1196} }

@ARTICLE{teng2016stochastic,
  author={Teng, Fei and Trovato, Vincenzo and Strbac, Goran},
  journal={IEEE Transactions on Power Systems}, 
  title={Stochastic Scheduling With Inertia-Dependent Fast Frequency Response Requirements}, 
  year={2016},
  volume={31},
  number={2},
  pages={1557-1566},
  doi={10.1109/TPWRS.2015.2434837}}

@article{wu2019method,
title = {A method to identify weak points of interconnection of renewable energy resources},
journal = {International Journal of Electrical Power \& Energy Systems},
volume = {110},
pages = {72-82},
year = {2019},
issn = {0142-0615},
doi = {https://doi.org/10.1016/j.ijepes.2019.03.003},
author = {Di Wu and Al Motasem Aldaoudeyeh and Milad Javadi and Feng Ma and Jin Tan and John N. Jiang}
}

@ARTICLE{teng2020synthetic,
  author={Chu, Zhongda and Markovic, Uros and Hug, Gabriela and Teng, Fei},
  journal={IEEE Transactions on Power Systems}, 
  title={Towards Optimal System Scheduling With Synthetic Inertia Provision From Wind Turbines}, 
  year={2020},
  volume={35},
  number={5},
  pages={4056-4066},
  doi={10.1109/TPWRS.2020.2985843}}

@misc{AER2019SAnews,
  title        = {AER approves ElectraNet spending on South Australia system strength},
  author       = {{Australian Energy Regulator}},
  year         = {2019},
  howpublished = {\url{https://tinyurl.com/3brr5452}}
}

@misc{Neutz2013STATCOMPQ,
  author       = {Michael Neutz},
  title        = {Power Quality: Voltage Stabilisation for Industrial Grids and Wind Farms with STATCOM},
  year         = {2013},
  organization = {ABB Switzerland Ltd.},
  howpublished = {\url{https://tinyurl.com/ybr5a596}} 
}

@misc{ABB2015SynconMining,
  title        = {Synchronous condensers in mining projects},
  author       = {{ABB}},
  year         = {2015},
  howpublished = {\url{https://tinyurl.com/54ud9837}},
}

@misc{AEMO2025RITTOandM,
  title        = {Victorian System Strength Requirement RIT-T PADR Webinar Slide Pack},
  author       = {{Australian Energy Market Operator (AEMO)}},
  year         = {2025},
  howpublished = {\url{https://tinyurl.com/3t95nyvb}},
}

@article{MEJIARUIZ2025110953,
title = {Multiple ancillary services provision by optimal control of aggregated inverter-based resources},
journal = {International Journal of Electrical Power \& Energy Systems},
volume = {171},
pages = {110953},
year = {2025},
issn = {0142-0615},
doi = {https://doi.org/10.1016/j.ijepes.2025.110953},
author = {Gabriel E. Mejia-Ruiz and Mario R. Arrieta Paternina and Zhihua Qu and Shehab Ahmed and Charalambos Konstantinou},
}






\end{document}